\documentclass[nofootinbib,prx,twocolumn,showpacs,superscriptaddress,notitlepage,superscriptaddress,amsmath]{revtex4-2}

\usepackage{lipsum}  
\usepackage{dsfont}
\usepackage{amsmath}
\usepackage{braket}
\usepackage{amssymb}
\usepackage{amsthm}
\usepackage{algpseudocode}
\usepackage{algorithm}
\usepackage{siunitx}
\usepackage{amsfonts}
\usepackage{comment}%https://www.overleaf.com/project/603d67a5abedda342ae4c069
\usepackage[normalem]{ulem}
\usepackage{graphicx}
\usepackage{color,framed}
\usepackage{hyperref}
\usepackage{enumerate}
\usepackage{lipsum}
\usepackage{slashed}
\usepackage{url}
\usepackage{bbm}
\usepackage{chngcntr}
\counterwithout{equation}{section}
\usepackage{tikz,pgfplots}
\usepackage{pgfplotstable}
\usepackage{siunitx}
\usepackage{comment}
\usepackage{graphicx}

\makeatletter
\newcommand{\setlabel}[1]{\edef\@currentlabel{#1}\label}
\makeatother

\usepgfplotslibrary{fillbetween}

\hypersetup{
    colorlinks=true, %set true if you want colored links
    linktoc=all,     %set to all if you want both sections and subsections linked
    linkcolor=blue,  %choose some color if you want links to stand out
}

\def \beq {\begin{equation}}
\def \eeq {\end{equation}}
\def \beqa {\begin{eqnarray}}
\def \eeqa {\end{eqnarray}}
\def \bseq {\begin{subequations}}
\def \eseq {\end{subequations}}

\newcommand \tr {{\rm tr}\,}

\newcommand{\sign}{{\rm sgn}\,}

\newcommand{\D}[1]{\text{d}#1}

\pgfplotsset{compat=1.18}
\begin{document}

\title{Dilution of error in digital Hamiltonian simulation}
\author{Etienne Granet}\author{Henrik Dreyer}
\affiliation{Quantinuum, Leopoldstrasse 180, 80804 Munich, Germany}
% \date{\today}

\begin{abstract}
We provide analytic, numerical and experimental evidence that the amount of noise in digital quantum simulation of local observables can be independent of system size in a number of situations.
We provide a microscopic explanation of this \emph{dilution of errors} based on the ``relevant string length" of operators, which is the length of Pauli strings in the operator at time $s$ that belong to the exponentially small subspace of strings that can give a non-zero expectation value at time $t$. We show that this explanation can predict when dilution of errors occurs and when it does not. We propose an error mitigation method whose efficiency relies on this mechanism. Our findings imply that digital quantum simulation with noisy devices is in appropriate cases scalable in the sense that gate errors do not need to be reduced linearly to simulate larger systems.
\end{abstract}

\maketitle

\section{Introduction}

It has often been assumed that, to obtain accurate results at the output of a quantum circuit with $t$ layers and $N$ gates per layer, each gate should have an infidelity of $1/(Nt)$ \cite{clinton2024towards,reiher2017elucidating,dalzell2023quantum}. In this scenario, a noisy quantum computer is expected to output states with global infidelity $\sim 1-1/e$ which bounds from above the relative error on all observables. Under this assumption, to tackle larger problems, quantum hardware makers would need to not only increase $N$, the number of qubits, but at the same time reduce the gate error as $1/N$. Since physical gate errors on current platforms are considered to be lower bounded by $10^{-4}$, one may conclude that useful quantum advantage cannot be achieved without error correction.

For global observables, as they appear for example in Shor's algorithm \cite{shor1999polynomial}, canonical phase estimation \cite{kitaev1995quantum}, modern filtering-based algorithms \cite{yang2022classical} or random circuit sampling \cite{arute2019quantum,decross2024computational}, this analysis is accurate. In contrast, in quantum many-body problems formulated in terms of local Hamiltonians, one is usually interested in the behavior of observables that are local (like magnetisation, particle or energy density, superconducting order parameters etc.). Such local observables are influenced only by a subset of gates inside of a light cone, which implies a loss of signal

\begin{align}
   \frac{\braket{O}_\mathrm{noisy}}{\braket{O}_\mathrm{noiseless}} = e^{- \varepsilon V}
    \label{eq_standard_noise_model}
\end{align}
for traceless observables $O$, where $V$ is number of operations in the light cone and $\varepsilon$ is proportional to the probability of failure for each of the operations. A more subtle argument replaces the circuit light cone with a smaller butterfly light cone which propagates with a buttefly velocity $v$ \cite{kechedzhi2024effective}. In $d$ dimensions, these lightcone arguments imply a loss of signal proportional to $\exp(-v^d t^{d+1})$ as long as $t v < \sqrt[d]{N}$ and $\exp(-Nt)$ afterwards, recovering the above-mentioned scaling issues for sufficiently deep circuits or high dimensions.

\begin{figure}[t!]
\begin{center}
\includegraphics[width=0.45 \textwidth]{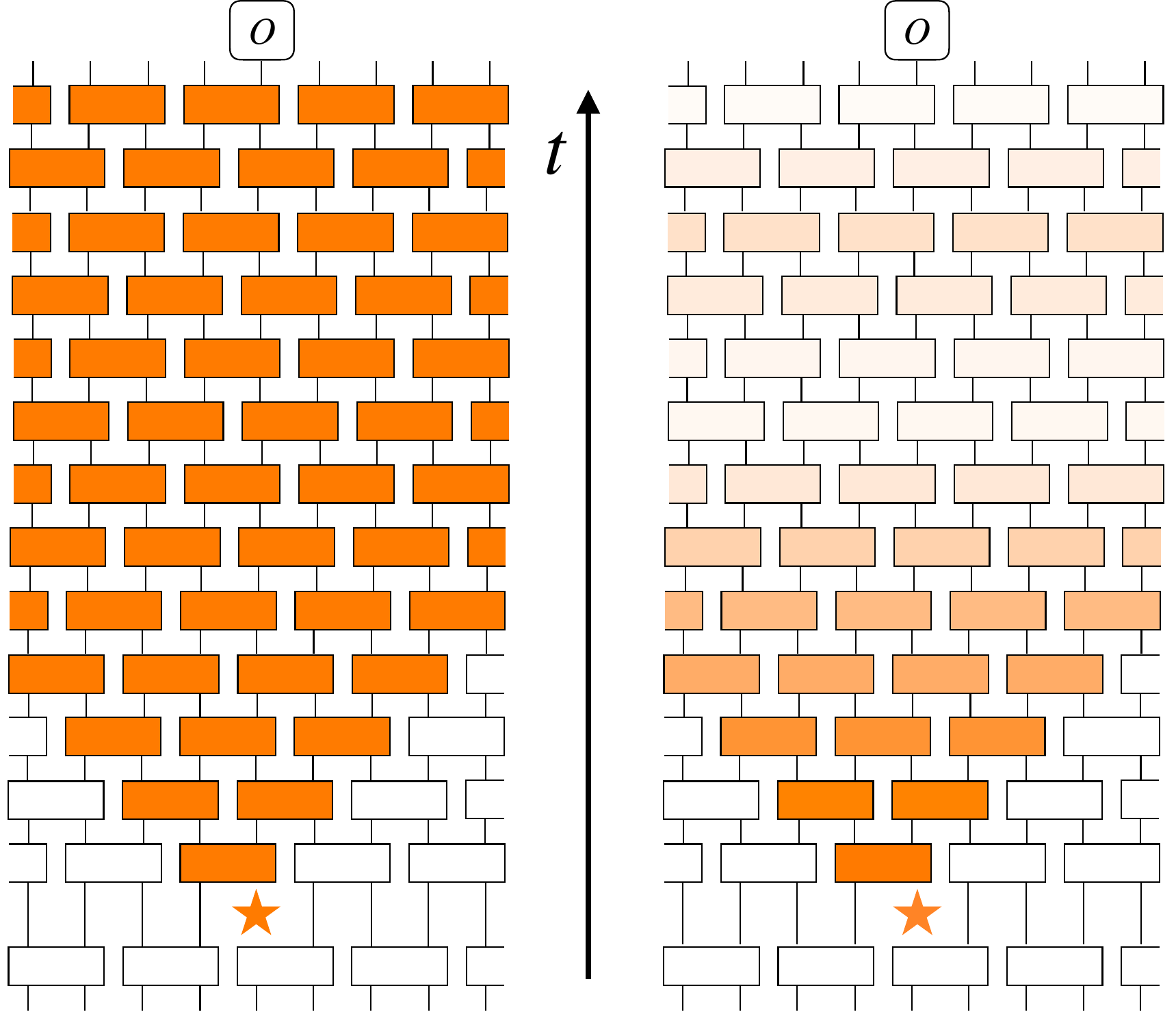}
\caption{\textbf{Dilution of Error.} In this paper, we consider the effect of gate errors (orange star) that occur during Trotterised Hamiltonian simulation on local observables $O$. Left: Usual light-cone arguments imply that errors spread with a finite velocity (which may be smaller than the one in the figure) but make no prediction on the magnitude of their effect within the cone. Right: We provide evidence that in certain settings errors dilute and their contribution oscillates around values that decrease with system size, often $\propto k/N$ where $k$ is the weight of the observable. 
}
\label{fig_conceptual}
\end{center}
\end{figure}

Here, we show that, in a variety of local Hamiltonian simulation tasks, all of these light cone arguments are too pessimistic. Instead, we propose a model in which errors dilute over time, i.e. 
\begin{align}
       \frac{\braket{O}_\mathrm{noisy}}{\braket{O}_\mathrm{noiseless}}  = e^{- \varepsilon \rho V}
       \label{eq_new_noise_model}
\end{align}
with a sensitivity to  errors $\rho$ that accounts for the fact that gate failures further in the past contribute less to the measurement of local observables in the present, because most errors get ``diluted" into a large subspace of operators with zero expectation value. 
Previous models correspond to the special case of uniform error density $\rho=1$. We provide evidence that the average density depends on the weight $k$ of the observable (i.e. its number of non-trivial Pauli matrices), in some cases linearly i.e. $\rho = kt/V$. In those cases, as long as $k=\mathcal{O}(1)$, the signal-to-noise ratio becomes independent of system size, of order $\mathcal{O}(N^0)$.
This implies that, above a problem-specific threshold, physical gate fidelities only need to increase mildly or not at all to reach quantum advantage when increasing the number of qubits in a given quantum computing architecture. 
Based on these findings, we develop and demonstrate two new error mitigation techniques, called \texttt{LIN} and \texttt{EXP}, both of which change the leading order error term $\varepsilon \rightarrow \varepsilon^2$ (and which can be extended to arbitrary higher order) with a low overhead in terms of shots.

There exist several related circumstances in quantum circuits where the effect of gate error or algorithmic error is smaller than expected. On analog quantum computers, low hardware noise has been observed and understood in some situations \cite{trivedi2022quantum,kashyap2024accuracy}. Noise scaling sub-linearly with system size has been observed for the return probability amplitude in controlled evolution \cite{dominguez2021decoherence}. 
Errors in Hamiltonian simulation incurred through Trotterization on digital quantum computers are well-known to be significantly smaller than standard bounds \cite{heyl2019quantum,zhao2022hamiltonian,zhao2024entanglement,tran2020destructive,zhao2021exploiting,childs2021theory}. In particular, in \cite{heyl2019quantum} Trotter error was found to be bounded with the number of qubits in certain cases. Noise or Trotter error was also sometimes found to have low impact on the performance of certain algorithms like adiabatic state preparation \cite{childs2001robustness,roland2005noise,kovalsky2023self,schiffer2024quantum,granet2024benchmarking,montanez2024towards}. 

The manuscript is organized as follows. In Section \ref{sec_experimental_evidence} we start with experimental evidence of dilution of errors in a quantum quench setup on Quantinuum's H1-1 quantum computer. In Section \ref{sec_microscopics} we propose a microscopic explanation for this effect based on the \emph{relevant string length} of the operator, which is loosely speaking the average length of Pauli strings in the time-evolved operator at time $s$ that can have a non-zero expectation value at time $t>s$. In Section \ref{sec_area_of_validity} we test the implications of this explanation and compare with numerical simulations, showing cases where dilution occurs and cases where it does not. Then in Section \ref{sec_error_mitigation} we propose two mitigation strategies with low overhead. In Section \ref{sec:freefermions} we show the dilution of errors in the exactly solvable 1D Ising model. Finally, in Section \ref{sec_trotter_error} we draw a connection between gate error and Trotter error. We give a summary and conclusion in Section \ref{sec_conclusion}.

\begin{figure*}[t!]
\begin{center}

\includegraphics[width=0.66 \textwidth]{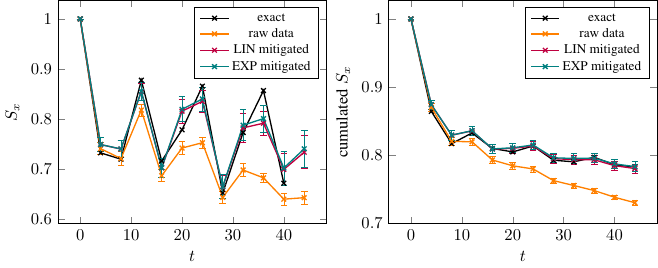}
\includegraphics[width=\textwidth]{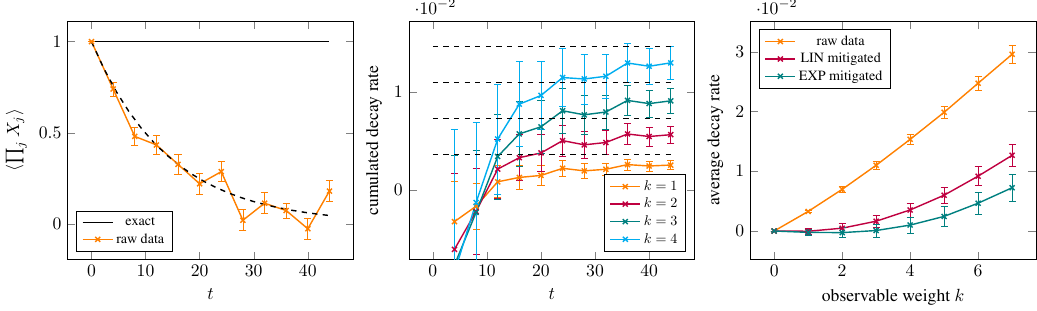}

\caption{\textbf{Experimental Data.} \emph{Top panels:} Expectation value of the magnetization $S_x^{(1)}$ as a function of time $t$ (instantaneous value in the left panel, and cumulated values in the right panel), raw data (orange), exact (black), mitigated values (teal and purple). The mitigation techniques are explained in Section \ref{sec_error_mitigation}. \emph{Bottom left panel:} Expectation value of the global observable $\prod_{j=1}^L X_j$ as a function of time $t$, raw data (orange) and exact (black). The dashed curve indicates a fit $e^{-0.07t}$. The noiseless value is constant due to symmetry. \emph{Bottom middle panel:} cumulated decay rate as a function of time $t$, for different observables $S_x^{(k)}$, computed from the raw H1-1 data. The dashed lines indicate the decay rate $\frac{k }{20}\times 0.07$. \emph{Bottom right panel:} Decay rate averaged over time for $t=16,...,44$ for the observable $S_x^{(k)}$, as a function of $k$, computed from the raw data (orange), \texttt{LIN} and \texttt{EXP} mitigated data (purple and teal), see Section \ref{sec_error_mitigation}.
}
\label{fig_decayrates_hardware}
\end{center}
\end{figure*}

\section{Experimental Evidence}
\label{sec_experimental_evidence}
We begin by describing an experiment carried out on the Quantinuum H1-1 trapped ion quantum computer \cite{chertkov2022holographic,pino2021demonstration} to assess the effect of realistic noise on observables of different weight. In particular, we simulate Trotterised quench dynamics of a transverse-field Ising model on an $N=5 \times 4$ square lattice with periodic boundary conditions. The quantum computer is initialised in the state $\ket{+}^{\otimes N}$ and quantum circuits $U^t$ are applied, where
\begin{align}\label{Utrotter}
    U = \left( \prod_{\braket{ij}} e^{-i \mathrm{d}t  Z_i Z_j} \right) \left( \prod_{i=1}^N e^{-i \mathrm{d}t \, h X_i} \right).
\end{align}
Here, the parameters are chosen as $h=3$, $\mathrm{d}t =0.15$ and $\braket{ij}$ refers to nearest neighbours on the square lattice. These values are close to the critical point of the 2D Ising model at zero temperature.
Each Trotter step consists of 40 two-qubit gates and their gate infidelity is known to dominate total error budgets. We detail this experimental setup in Appendix \ref{appendix-experiemnt}. After $t$ Trotter steps, the wavefunction is collapsed in the local $X$-basis, giving us access to the observables
\begin{align} \nonumber
    S_x^{(1)} &= \frac{1}{N} \sum_{i=1}^N X_i \\
    S_x^{(2)} &= \frac{1}{\binom{N}{2} } \sum_{i \neq j} X_i X_j \\ \nonumber
    \vdots \\
    S_x^{(N)} &= X_1X_2...X_N\,.
    \label{eq_definition_Sxk}
\end{align}
Evidently, $S_x^{(k)}$ has weight $k$ and interpolates between a local observable for $k=1$ and a global observable for $k=N$. 
In Fig.~\ref{fig_decayrates_hardware}, we show experimental data for the magnetization $S_x^{(1)}$ and for the global parity $S_x^{(N)}$ at times $t= 4, 8, 12, \dots, 44$. Focusing first on the global parity, we see that the signal-to-noise ratio $\braket{S_x^{(N)}}_\mathrm{noisy} / \braket{S_x^{(N)}}_\mathrm{noiseless} \approx \exp(-0.07 t)$ decays exponentially and the exponent is compatible with equation~(\ref{eq_standard_noise_model}) and an effective two-qubit gate fidelity $F = \exp(-\varepsilon) = \exp(-0.07/40) \approx 0.998$, which is in line with component benchmark data \cite{datasheet}. However, the experimental raw data for the magnetization $S_x^{(1)}$ is significantly less attenuated, with the cumulated magnetization $\frac{4}{t}\sum_{s=1}^{t/4} \langle S_x^{(1)}(4s)\rangle$ conserving around $90\%$ of the exact value at $t=40$, compared to $6\%$ for the global parity.

To investigate this further, we study now the behaviour of observables $S_x^{(k)}$ with $1<k<N$. We define the decay rate for a given observable $O$ as
\begin{align}
\label{eq_noise_rate}
    \lambda(t) := -\frac{1}{t} \ln \frac{\braket{O(t)}_\mathrm{noisy}}{\braket{O(t)}_\mathrm{noiseless}}.
\end{align}
According to the standard light cone model~(\ref{eq_standard_noise_model}) with velocity $v$, in this two-dimensional system, one would expect an initial ramp $\lambda(t) \sim \epsilon t^2 v k$ at very early times followed by a regime when the light cone hits the boundary of the system, upon which $\lambda(t) = \epsilon N - \mathcal{O}(1/t)$. Crucially, at large times, standard light cone arguments predict a constant decay rate $\epsilon N$ for any observable and, in our particular setting, would imply a decay rate constant in $k$ for the different $S_x^{(k)}$. %, cf.~(\ref{eq_definition_Sxk}). 
The experimental data, shown in the middle bottom panel of Fig.~\ref{fig_decayrates_hardware} is in stark contrast with this prediction. Instead, at late times, the decay rate $r$ depends linearly on the weight $k$ of the observable.

A minimal generalisation of light cone theory can account for this observation: By allowing the possibility of an error inside the light cone to affect the final measurement with average probability $\rho = k/N \leq 1$, and replacing eq.~(\ref{eq_standard_noise_model}) with~(\ref{eq_new_noise_model}), we obtain a model that is compatible with all experimental data for $k=1, \dots N$. Proposing a physical explanation for the origin of this new parameter is what we turn to next.

\section{Microscopic Origin for the Dilution of Error
\label{sec_microscopics}}
\subsection{Definition of the $\Sigma$ expansion}\label{sigmasection}

In the previous section, we have provided experimental data to refute a standard light cone model of errors in Trotterized local Hamiltonian simulation and introduced an ad-hoc parameter $\rho$ to remedy the situation. The purpose of the present section is to explain the origin of this parameter and bestow it with some physical intuition.

We consider quantum circuits on $N$ qubits that can be decomposed into layers $U = U(T) \dots U(2) U(1)$ and denote by $U_{t,s} = U(t) \dots U(s+2) U(s+1)$ subcircuits implementing layers from $s+1$ to $t$ when $s< t$, and $U_{s,t}=U_{t,s}^\dagger$. We will also write $O(t)=U_{t,0}O U_{t,0}^\dagger$ for $O$ an observable. Further, we assume that the noise can be modelled by the application of a quantum channel $\mathcal{N}$ after each layer.
For simplicity and lightness of notations, we will work with a noise channel including only one type of error occurring on single qubits, written as
\begin{equation}\label{noisechannel}
\begin{aligned}
     &\mathcal{N}(\rho)=\mathcal{N}_1(...\mathcal{N}_N(\rho)...)\,,\\
     &\text{with }\qquad\mathcal{N}_j(\rho)=(1-\varepsilon)\rho+\varepsilon K_{j} \rho K_{j}^\dagger\,,
\end{aligned}
\end{equation}
with $0<\varepsilon<1$ the probability of error and $K_{j}$ a Pauli matrix located at site $j$.
Generalization to multiple Kraus operators or to operators acting on multiple qubits is straightforward. In particular we will often consider the case of three Kraus operators $K_{j;1}=X_j/\sqrt{3}$, $K_{j;2}=Y_j/\sqrt{3}$, $K_{j;3}=Z_j/\sqrt{3}$, implementing a one-qubit depolarizing channel \cite{nielsen2010quantum}. One can decompose the effect of the noise on the final measurement into sectors with a fixed number of errors:
\begin{align}\label{eq_noisy}
    &\langle O(T)\rangle_{\rm noisy}=\sum_{n=0}^{TN}\varepsilon^n (1-\varepsilon)^{TN-n}D_n\,,
\end{align}
with $D_n$ the sector with exactly $n$ errors, e.g., 
\begin{align}\label{definitiond}
D_0 &= \braket{O(T)} \nonumber\\ 
D_1 &= \sum_{j_1,s_1} \braket{U_{T,s_1} K_{j_1} U_{s_1,0} O U_{s_1,0}^\dagger K_{j_1}^\dagger U_{T,s_1}^\dagger} \\
D_2 &= \sum_{\substack{j_1,s_1\\j_2,s_2\\ \text{ordered}}}
\braket{ K_{(2)} K_{(1)} O K_{(1)}^\dagger K_{(2)}^\dagger } \,,\nonumber
\end{align}
where $\langle \cdot \rangle$ denotes the (noiseless) expectation value within a fixed initial state. We used the simplified notation $K_{(i)}=U_{s_{i+1},s_{i}} K_{j_1} U_{s_i,s_{i-1}}$ with $s_0=0,s_{n+1}=T$. Here, ``ordered" means that the sites and times are ordered as $1\leq s_1\leq s_2\leq ...\leq T$, and if $s_1=s_2$ then $j_1<j_2$. This expression might suggest that the noise attenuates the noiseless expectation value $D_0=\braket{O(T)}$ by a factor $(1-\varepsilon)^{TN}$. However, we saw in Section~\ref{sec_experimental_evidence}  that the signal can be much less attenuated for local observables. This means that the remaining terms in the right-hand side of~\eqref{eq_noisy} are in good part close to the noiseless expectation value, and that the expansion \eqref{eq_noisy} in terms of the number of errors is misleading.

A much more useful expression appears when we regroup~\eqref{eq_noisy} in orders of $\varepsilon$:
\begin{equation}\label{decomposition}
    \begin{aligned}
        \langle O(T)\rangle_{\rm noisy}=\sum_{n=0}^{TN}\varepsilon^n \langle\Sigma_n(T)\rangle
    \end{aligned}
\end{equation}
where we have defined the operators
\begin{equation}
\begin{aligned}
    \Sigma_0(T)&=O(T)\\
\Sigma_1(T)&=\sum_{j_1,s_1} U_{T,s_1}[K_{j_1},U_{s_1,0}O U_{s_1,0}^\dagger]K_{j_1}^\dagger U_{T,s_1}^\dagger\\
\Sigma_2(T)&=\sum_{\substack{j_1,s_1\\j_2,s_2\\ \text{ordered}}} [K_{(2)},[K_{(1)},O] K^\dagger_{(1)}]K^\dagger_{(2)}\, ,
\end{aligned}
\end{equation}
and so on for higher orders. At small error rate $\varepsilon$, the signal-to-noise ratio in \eqref{eq_new_noise_model} is
\begin{equation}
    \frac{\langle O(t)\rangle_{\rm noisy}}{\langle O(t)\rangle_{\rm noiseless}}=1+ \epsilon \frac{\langle \Sigma_1(t)\rangle}{\langle \Sigma_0(t)\rangle}+\mathcal{O}(\varepsilon^2)\,.
\end{equation}
In this regime, the dilution factor $\rho$ can thus be expressed as
\begin{equation}\label{rhosigma}
    \rho=-\frac{1}{NT}\frac{\langle \Sigma_1(t)\rangle}{\langle \Sigma_0(t)\rangle}+\mathcal{O}(\varepsilon)\,.
\end{equation}

\subsection{Relevant Pauli string length mechanism}
\subsubsection{Noise and Pauli string length}
Let us focus on the term $\Sigma_1(t)$, initial state $|+...+\rangle$ and consider the insertion of errors at time $s$. We decompose the observable at that time as
\begin{equation}\label{paulistring}
    O(s)=\sum_{P\in \mathcal{P}}c_P(s)P\,,
\end{equation}
where 
\begin{equation}
    \mathcal{P}=\{M_1\otimes...\otimes M_N\,,\, M_j=I,X,Y,Z\}
\end{equation}
denotes the set of all $4^N$ Pauli strings, and with $c_P(s)$ real coefficients. We have $\frac{1}{2^N}\tr[O(s)^\dagger O(s)]= \sum_{P\in\mathcal{P}}|c_P(s)|^2$, and this is independent of $s$ due to unitarity of the noiseless evolution. 

The effect of the commutator with $K_j$ is particularly simple in this decomposition. For a Pauli string $P$, we have $[K_j,P]K_j=0$ if $K_j$ commutes with $P$, and $[K_j,P]K_j=-2P$ if $K_j$ anticommutes with $P$. In particular, no error at site $j$ can have an effect on the Pauli string $P$ if the Pauli matrix in $P$ at site $j$ is $I$. When summed over $j$, the total influence of noise can thus be bounded by the \emph{length} of the Pauli string as follows
\begin{equation}\label{kpkalpha}
    \sum_{j=1}^N [K_j,P]K_j=-2\alpha P\,,\qquad 0\leq \alpha\leq \ell_P\,,
\end{equation}
where $\ell_P$ is the number of Pauli matrices in the Pauli string $P$ that are different from $I$. We will define as well the linear operator $\ell$ by $\ell(P)=\ell_P P$. In case of a depolarizing channel, we have
\begin{equation}
    \frac{1}{3}\sum_{K\in\{X,Y,Z\}}\sum_{j=1}^N [K_j,P]K_j=-\frac{4\ell_P}{3} P\,.
\end{equation}
We thus see that the structure of the coefficients $c_P$ and their value according to the length of the string $P$ are key to understanding the effect of noise. For example, if all the string lengths were bounded in \eqref{paulistring}, namely $c_P(s)=0$ whenever $\ell_P>\ell_{\rm max}$ for some $\ell_{\rm max}$, then we would have $|\langle \Sigma_1(T)\rangle |\leq 2\ell_{\rm max} T$ independently of system size $N$. The typical behaviour of Pauli string length in quantum circuits is however very defavorable to this argument. Defining the average string length as
\begin{equation}
\begin{aligned}
 \mathcal{L}(s)\equiv\frac{\tr[O(s)^\dagger \ell(O(s))]}{\tr[O(s)^\dagger O(s)]}=\frac{\sum_{P\in\mathcal{P}}\ell_P|c_P(s)|^2}{\sum_{P\in\mathcal{P}}|c_P(s)|^2}\,,
\end{aligned}
\end{equation}
it is well-known that $\mathcal{L}(s)$ typically grows linearly with time $s$ or circuit depth, both in random circuits and in Hamiltonian simulation, up to saturating at some value $\mathcal{L}(s)=\mathcal{O}(N)$ \cite{nahum2018operator,schuster2023operator,chan2018solution,von2018operator,khemani2018operator,parker2019universal}. For an equal superposition of all $4^N$ Pauli strings, this average length would be $3N/4$. This behaviour is also expected from a simulation complexity perspective: if the string length remained bounded $\mathcal{O}(N^0)$, one could simulate the system in time that is polynomial in system size by keeping track of a number of non-zero Pauli strings that is polynomial in $N$ \cite{schuster2024polynomial}. The coefficient $\alpha$ in \eqref{kpkalpha} obtained from this simple ``Pauli matrix counting" argument cannot thus be the noise rate, as it would lead to a noise $\mathcal{O}(N)$ per time step once the light cone has reached the boundaries, which is incompatible with experimental and numerical observation.

\subsubsection{Pauli strings and expectation values}
The length of strings in the decomposition \eqref{paulistring} determines the influence of noise \emph{on the operator} $O(t)$. However, the expectation value of the operator within a state only depends on part of the Pauli strings in \eqref{paulistring}. Indeed, a Pauli string $P$ evaluated in the initial state $|+...+\rangle$ vanishes $\langle +...+|P|+...+\rangle=0$ whenever $P$ contains a $Y$ or $Z$. Hence, when taking the expectation value at time $t$, we have
\begin{equation}
\begin{aligned}
    \langle +...+|O(t)|+...+\rangle&=\sum_{P\in \mathcal{P}_x} c_P(t)\langle +...+|P|+...+\rangle\\
    &=\sum_{P\in \mathcal{P}_x} c_P(t)\,,
\end{aligned}
\end{equation}
where $\mathcal{P}_x=\{M_1\otimes...\otimes M_N\,, M_j=I,X\}$ denotes the set of all $2^N$ Pauli strings in $X$ only. We thus see that among all the $4^N$ possible Pauli strings, only a tiny fraction $1/2^N$ of all strings contributes to the expectation value at time $t$. Most of the information contained in the Pauli string decomposition of $O(t)$ -- as well as the scrambling of this information by noise -- is thus \emph{irrelevant} to the result of a measurement. The typical ``relevance" of a string $P$ is related to its length: If one draws at random a Pauli string $P$ of length $k$, there is a probability $1/3^k$ that it is relevant and contributes to the expectation value, since there are $3^k {N\choose k}$ strings of length $k$ involving any Pauli matrices, but only ${N\choose k}$ strings of length $k$ involving just $I$ and $X$. 

This observation is key to the \emph{dilution of errors} presented in this paper. Although a string $P$ drawn at random will be more likely affected by an error $K_j$ if the length $\ell_P$ is large, it will modify the expectation value only with an exponentially small probability $1/3^{\ell_P}$. Errors on large strings are \emph{diluted} into an enormous state space most of which does not affect the expectation value of the observable. Only errors on small strings contribute to the noise since they are much more likely of being relevant to the expectation value.

\begin{figure*}
\begin{center}

\includegraphics[width=0.66 \textwidth]{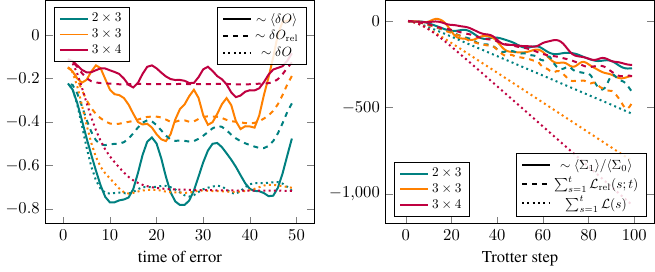}

\caption{\emph{Left panel:}  Comparison of $\frac{\langle U_{t,s}\delta O U_{t,s}^\dagger\rangle}{\langle O(t)\rangle}$ (solid lines), $r(s;t)=\frac{\tr[O_{\rm rel}(t)\delta O_{\rm rel}]}{\tr[O_{\rm rel}(t)O_{\rm rel}(t)]}$ (dashed lines) and $\frac{\tr[O(t)\delta O(t)]}{\tr[O(t)O(t)]}$ (dotted lines), all for $\delta O=\sum_{K=X,Y,Z}[K_1,O(s)]K_1$, as a function of time of error $s$, for total time $t=50$, in the same setting as in Section \ref{sec_experimental_evidence} with $h_x=1$ and ${\rm d}t=0.1$, for the observable $S_x$, for different system sizes. \emph{Right panel:} Same as left panel but with noise on all qubits, and for $2 {\rm cum}(\langle \Sigma_1(t)\rangle)/{\rm cum}(\langle \Sigma_0(t)\rangle)$ (solid lines), relevant length $-\frac{4}{3}\sum_{s=1}^t \mathcal{L}_{\rm rel}(s;t)$ (dashed lines) and total length $-\frac{4}{3}\sum_{s=1}^t \mathcal{L}(s)$ (dotted lines), as a function of time $t$. We defined the cumulated function ${\rm cum}(f(t))=\sum_{s=1}^t f(s)$.}
\label {compare}
\end{center}
\end {figure*}

\subsubsection{Relevant strings}
\label{relation_relevant_noise}
Let us now formalize this intuition. We consider an error $K_j$ that occurs at time $s$ in the observable $O(t)$. We define the \emph{relevant subspace} of strings at time $s$ for time $t$ as
\begin{equation}
    S_{\rm rel}(s;t)={\rm span}\{ U^\dagger_{t,s}P U_{t,s}\,,\quad P\in \mathcal{P}_x\}
\end{equation}
where ${\rm span }(S)$ denotes the space generated by all linear combinations of the set $S$. We note that the Pauli strings $P$ in this definition are evolved backward in time. This is a subspace of dimension $2^N$ in the space of operators that is of dimension $4^N$. It corresponds exactly to the subspace of strings at time $s$ that can contribute to the expectation value at time $t$. We then define $O_{\rm rel}(s;t)$ as the projection of $O(s)$ onto $S_{\rm rel}(s;t)$, with respect to the trace inner product $\tr[O^\dagger O']$ for $O,O'$ operators. This operator is the part of $O(s)$ that contributes to the expectation value at time $t$. The explicit expression for this projection is
\begin{equation}
    O_{\rm rel}(s;t)=\frac{1}{2^N}\sum_{P\in \mathcal{P}_x} U^\dagger_{t,s}P U_{t,s} \,\tr[U^\dagger_{t,s}PU_{t,s} O(s)]\,.
\end{equation}
It can be rewritten exactly as
\begin{equation}
    O_{\rm rel}(s;t)=U_{t,s}^\dagger\Pi_X (U_{t,0} O U_{t,0}^\dagger) U_{t,s}\,,
\end{equation}
where $\Pi_X$ denotes the superoperator that projects onto $X$ strings only, given  by
\begin{equation}
    \Pi_X(\rho)=\frac{1}{2^N}\sum_{n_1,...,n_N\in \{0,1\}}X_j^{n_j}\rho X_j^{n_j}\,.
\end{equation}
By construction, the relevant space $S_{\rm rel}(s;t)$ satisfies the following property: for any operator $\delta O$ that is orthogonal to $S_{\rm rel}(s;t)$, perturbing $O(s)$ by $\delta O$ does not modify the expectation value at time $t$
\begin{equation}\label{deltaO}
    \langle U_{t,s}(O(s)+\delta O)U^\dagger_{t,s}\rangle=\langle O(t)\rangle\,.
\end{equation}
In our case, we are interested in the perturbation due to noise as appearing in the $\Sigma$ expansion
\begin{equation}
    \delta O= \sum_{j=1}^N[K_j,O(s)]K_j\,.
\end{equation}
As only the projection of $\delta O$ onto $S_{\rm rel}(s;t)$ can modify the expectation value of the observable, we give it a name $(\delta O)_{\rm rel}(s;t)$, that is
\begin{equation}
    (\delta O)_{\rm rel}(s;t)=\sum_{j=1}^NU_{t,s}^\dagger\Pi_X (U_{t,s} [K_j,O(s)]K_j U_{t,s}^\dagger) U_{t,s}\,.
\end{equation}
The noisy expectation value at time $t$ with this perturbation at time $s$ is thus $\langle U_{t,s}O_{\rm rel}(s;t)U^\dagger_{t,s} \rangle+\langle U_{t,s}(\delta O)_{\rm rel}(s;t)U^\dagger_{t,s} \rangle$. 

Being in the relevant space is a necessary condition for the error to modify the expectation value, but it does not inform us about the amplitude of the modification. In the case where $(\delta O)_{\rm rel}(s;t) \propto O_{\rm rel}(s;t)$, the perturbation on the expectation value is proportional to the noiseless expectation value, and the inner product $\tr[O_{\rm rel}(s;t)^\dagger (\delta O)_{\rm rel}(s;t)]$ divided by $\tr[O_{\rm rel}(s;t)^\dagger O_{\rm rel}(s;t)]$ provides us exactly with the proportionality factor. We thus define
\begin{equation}
\begin{aligned}
    r(s;t)&\equiv\frac{\tr[O_{\rm rel}(s;t)^\dagger (\delta O)_{\rm rel}(s;t)]}{\tr[O_{\rm rel}(s;t)^\dagger O_{\rm rel}(s;t)]}\\
    &=\frac{\tr[O_{\rm rel}(s;t)^\dagger \delta O(s)]}{\tr[O_{\rm rel}(s;t)^\dagger O(s)]}\,,
\end{aligned}
\end{equation}
as an estimate of the influence of the error at time $s$ on the expectation value at time $t$, in the sense that we expect
\begin{equation}
    \langle O\rangle_{\rm noisy}\approx \langle O\rangle_{\rm noiseless}(1+r(s;t))\,.
\end{equation}
In the case of a sum of one-qubit depolarizing channels applied on each qubit, the perturbation $\delta O$ would be proportional to $\ell(O(s))$. This leads us to define the \emph{relevant string length} at time $s$ for time $t$ as
\begin{equation}
   \mathcal{L}_{\rm rel}(s;t)=\frac{\tr[O_{\rm rel}(s;t)^\dagger \ell(O(s))]}{\tr[O_{\rm rel}(s;t)^\dagger O(s)]}\,.
\end{equation}
Intuitively, this quantity gives an estimate of the average length of the strings in the operator $O(s)$ that contribute to the expectation value at time $t$. However,  this quantity can in principle be negative or larger than $N$.  We postulate that $\mathcal{L}_{\rm rel}(s;t)$ is a good measure of the effect of depolarizing noise at time $s$ on the expectation value at time $t$. Namely, that the dilution factor in \eqref{eq_new_noise_model} is approximately 
\begin{equation}\label{rholrel}
    \rho\approx \frac{C}{N t}\sum_{s=1}^t\mathcal{L}_{\rm rel}(s;t)\,,
\end{equation}
with a prefactor $C$ of order $\mathcal{O}(1)$ that depends on the specific noise model.
To support this claim, we provide in the left panel of Figure \ref{compare} numerical simulations comparing $\frac{\langle U_{t,s}\delta O U_{t,s}^\dagger\rangle}{\langle O(t)\rangle}$ with $r(s;t)$, where the perturbation $\delta O$ is obtained from a depolarizing channel on one qubit. As a baseline comparison, we also plot $\frac{\tr[O(t)\delta O(t)]}{\tr[O(t)O(t)]}$, i.e. the same quantity as $r(s;t)$ but without projecting onto the relevant space. We observe that $r(s;t)$ and the perturbation on the expectation values are indeed of the same order, and in particular decreasing with $N$, whereas $\frac{\tr[O(t)\delta O(t)]}{\tr[O(t)O(t)]}$ remains of order $\mathcal{O}(N^0)$. Then in the right panel, we compare $\sum_{s=1}^t \mathcal{L}_{\rm rel}(s;t)$ with the value of $\langle \Sigma_1(t)\rangle/\langle \Sigma_0(t)\rangle$ which is related to the sensitivity to errors at small error rate through \eqref{rhosigma}. We also provide the value of $\sum_{s=1}^t \mathcal{L}(s)$ for baseline comparison, i.e. the sum of the total length at each time point. We observe that $\langle \Sigma_1(t)\rangle/\langle \Sigma_0(t)\rangle$ and the relevant length are indeed of the same order. In particular, while both grow linearly with $t$ as expected, they are seen to be of order $\mathcal{O}(N^0)$ with respect to system size, contrary to the a priori expectation $\mathcal{O}(N)$. In contrast, the total cumulated length $\sum_{s=1}^t \mathcal{L}(s)$ is seen to be of order $\mathcal{O}(tN)$.

Besides this numerical evidence, we will present in Section \ref{toymodelrelation} below a toy model in which the relation \eqref{rholrel} between $\rho$ and $\mathcal{L}_{\rm rel}(s;t)$ can be explicitly derived.

\begin{figure*}
\begin{center}

\includegraphics[width=\textwidth]{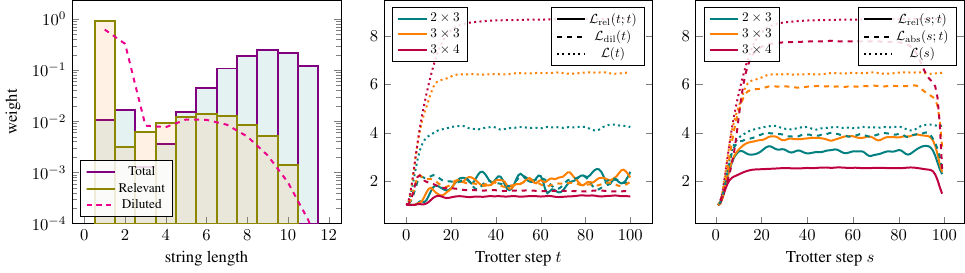}

\caption{\emph{Left panel:} Histogram of string length in $X_1(t)$ after $100$ Trotter steps in the setup of Section \ref{sec_experimental_evidence} with $h_x=1$, ${\rm d}t=0.1$ and system size $3\times 4$, taking into all Pauli matrix type (Total), only $X$ strings (Relevant), and all Pauli matrix type but with a dilution factor $1/3^k$ for string length $k$ (Diluted). The histograms are normalized to $1$ for each of the three cases. \emph{Middle panel:} In the same setting as left panel, evolution of the average string lengths as a function of Trotter step: $\mathcal{L}(t)$ (Total), $\mathcal{L}_{\rm rel}(t;t)$ (Relevant) and $\mathcal{L}_{\rm dil}(t)$ (Diluted). \emph{Right panel:} $\mathcal{L}_{\rm rel}(s;t)$ (solid), $\mathcal{L}_{\rm abs}(s;t)$ (dashed) and $\mathcal{L}(s)$ (dotted) as a function of $s$, for $t=100$, for different system sizes.}
\label {systemsizerelevant}
\end{center}
\end {figure*}

\subsubsection{System-size behaviour of the relevant string length at time $t$ for time $t$}
\label{revelantlengthtt}

We now wish to understand the behaviour of the relevant string length with system size. To that end, we need to recall a few results on the length distribution of Pauli strings in Hamiltonian simulation.

As stated at the beginning of this Section, the growth of the string length $\mathcal{L}(t)$ with time, up to saturation at $\mathcal{O}(N)$, is a typical situation for quantum circuits. In absence of conservation laws, for deep circuits the length distribution becomes concentrated at $\mathcal{O}(N)$ lengths, and the weight of short strings $\mathcal{O}(1)$ becomes exponentially small in system size \cite{nahum2018operator,schuster2023operator,chan2018solution,von2018operator,khemani2018operator,parker2019universal}. However, in presence of conserved quantities, short strings can maintain a non-zero weight even for deep circuits \cite{khemani2018operator,schuster2023operator}. Let us take for example the case of Hamiltonian simulation and neglect Trotter error. Then the Hamiltonian $H$ commutes with the time evolution operator $e^{iHt}$, so that
\begin{equation}
    \tr[H O(t)]=\tr[H O(0)]\,.
\end{equation}
We consider the Ising model of Section \ref{sec_experimental_evidence} with $O=S_x$, and denote $c_X(t)$, $c_{ZZ}(t)$ the coefficients of Pauli strings $X_j$ and $Z_jZ_k$ in $O(t)$ for $j,k$ neighbours on the lattice, assumed for simplicity to be independent of lattice site (as for example it is the case for a square lattice with periodic boundary conditions). Then, from one Trotter step to the other, neglecting Trotter error we must have
\begin{equation}\label{conservationlaw}
    c_X(t)h+2c_{ZZ}(t)=\frac{h}{N}\,,
\end{equation}
the right-hand side being the value of the left-hand side at $t=0$. We note that this is \emph{not} an energy conservation equation for the state, since these weights are those of the observable $O(t)$ and not of the state. We will see below in Section \ref{sec_time_dep} that a similar equation holds for time-dependent settings where no energy is conserved. This equation ensures that if the observable we measure has a non-zero overlap with the Hamiltonian, then some short strings have a non-exponentially-small weight that is preserved under the time evolution. Indeed, since the right-hand side is non-zero, $c_X(t)$ and $c_{ZZ}(t)$ cannot both become arbitrary small. We will thus refer to this kind of equation as a \emph{string transfer equation} (STE), to convey the idea that the weight of short strings is (potentially partially) transferred from one step to the other instead of being diluted into longer strings. We use a more general terminology than ``conservation" to anticipate the time-dependent case further in the manuscript. Similarly, from the commutation of $H^2$ with $e^{iHt}$ one obtains STEs for some longer strings. We note that these STEs however do not preclude most of the string weight to be concentrated on long strings, and the average length still saturates at $\mathcal{O}(N)$ quickly as time grows.

Let us now investigate how this picture generalizes to the relevant string length $\mathcal{L}_{\rm rel}(s;t)$. We first start with the case $s=t$, namely the average length at time $t$ of strings that are composed \emph{only} of a single type of Pauli matrices, for example $X$
\begin{equation}
    \mathcal{L}_{\rm rel}(t;t)=\frac{\sum_{P\in\mathcal{P}_x}|c_P(t)|^2\ell_P}{\sum_{P\in\mathcal{P}_x}|c_P(t)|^2}\,.
\end{equation}
For generic quantum circuits, we expect the weight of strings to be fairly uniformly distributed among different Pauli matrix type $X,Y,Z$. As a consequence, among strings of length $k$, the weight of Pauli strings only composed of $X$ matrices should contribute only with a proportion $1/3^k$, since there are ${N\choose k}3^k$ generic strings of length $k$, but only ${N\choose k}$ composed only of $X$. This suggests thus the following relation
\begin{equation}
    \mathcal{L}_{\rm rel}(t;t)\approx \frac{\sum_{P\in\mathcal{P}}|c_P(t)|^2\ell_P 3^{-\ell_P}}{\sum_{P\in\mathcal{P}}|c_P(t)|^2 3^{-\ell_P}}\equiv \mathcal{L}_{\rm dil}(t)\,.
\end{equation}
The interpretation is that $X$-strings become ``diluted" in the enormous space of operators as their length grows, which will favor short strings. We thus expect that, as long there are some short strings with a weight that is not exponentially small, the relevant string $\mathcal{L}_{\rm rel}(t;t)$ at time $t$ for time $t$ should remain of order $\mathcal{O}(N^0)$. For generic circuits without STE, we thus should have $\mathcal{L}_{\rm rel}(t;t)=\mathcal{O}(N)$ after some time $t$, whereas in case of an STE, this implies that we should have $\mathcal{L}_{\rm rel}(t;t)=\mathcal{O}(N^0)$ for all times.  

To make this intuition more quantitative, let us denote 
\begin{equation}
    p_k=\sum_{\substack{P\in\mathcal{P}\\ \ell_P=k}}|c_P|^2\,,
\end{equation}
the probability that a string in $O(t)$ taken at random has length $k$. The dilution roughly modifies this probability distribution into $p_k 3^{-k}$, up to a global normalization factor. In this new distribution, short strings of length e.g. $k=1$ dominate if
\begin{equation}\label{conditionp1}
    p_1\gg \sum_{k=2}^N p_k 3^{-k}\,.
\end{equation}
Let us take an example where all strings of length $k\neq 1$ have same weight, and take $p_1$ as a free parameter. Then, due to the constraint $\sum_{k=0}^Np_k=1$, we have for $k\neq 1$
\begin{equation}\label{pk}
    p_k=\frac{1-p_1}{1-3N/4^N} \frac{3^k}{4^N}{N\choose k}\,.
\end{equation}
The average string length is
\begin{equation}
\begin{aligned}
    \sum_{k=0}^N kp_k &=p_1+\frac{1-p_1}{1-3N/4^N} \left( 
 \frac{3N}{4} - \frac{3N}{4^N} \right)\\
 &\sim (1-p_1)\frac{3N}{4}\,,
\end{aligned}
\end{equation}
where by $\sim x$ we mean $x+o(x)$ as $N\to\infty$. This is $\mathcal{O}(N)$ for any $p_1<1$. On the other hand, we have
\begin{equation}
\begin{aligned}
    \sum_{k=0}^N p_k 3^{-k} &=\frac{p_1}{3}+\frac{1-p_1}{1-3N/4^N} \left( 
 \frac{1}{2^{N}} - \frac{N}{4^N} \right)\\
 &\sim \frac{p_1}{3}+\frac{1-p_1}{2^{N}}\,,
\end{aligned}
\end{equation}
which is $\sim \frac{p_1}{3}$ if $p_1\gg \frac{1}{2^N}$, and $\sim \frac{1}{2^N}$ otherwise. Finally we have
\begin{equation}
\begin{aligned}
    \sum_{k=0}^N kp_k 3^{-k} &=\frac{p_1}{3}+\frac{1-p_1}{1-3N/4^N} \left( 
 \frac{N}{2^{N+1}} - \frac{N}{4^N} \right)\\
 &\sim \frac{p_1}{3}+(1-p_1)\frac{N}{2^{N+1}}\,,
\end{aligned}
\end{equation}
which is $\sim \frac{p_1}{3}$ if $p_1\gg \frac{N}{2^N}$, and $\sim \frac{N}{2^{N+1}}$ otherwise. The relevant string length
\begin{equation}
    \mathcal{L}_{\rm rel}=\frac{\sum_{k=0}^N kp_k 3^{-k}}{\sum_{k=0}^N p_k 3^{-k}}
\end{equation}
behaves thus as $\sim N/2$ if $p_1\ll \frac{1}{2^N}$, and as $\sim 1$ if $p_1\gg \frac{N}{2^N}$. In absence of STE, if the weight of Pauli strings is uniform, we would have $p_1=\frac{3N}{4^N}$ and so a relevant string length becomes $\mathcal{O}(N)$. However, in case of a STE that constrains for example $p_1=\frac{1}{N}$, we have a relevant string length $\mathcal{O}(1)$ despite a total string length $\mathcal{O}(N)$.

We present in the middle and right panels of Figure \ref{systemsizerelevant} numerical evidence for these behaviours in the case of Hamiltonian simulation of an Ising model. In the left panel, we show a histogram of the weight of strings in $S_x(t)$ at fixed time $t$ as a function of their length, taking into account all types of strings, or only $X$-strings, namely the sum of $|c_P(t)|^2$ at fixed $\ell_P$, with or without the constraint that $P$ is a $X$ string. We see that for generic strings, the weight is strongly concentrated at large length, with a non-negligible weight at short length due to the STE coming from the non-zero trace overlap between the observable $S_x$ and the Hamiltonian $H$. However, for relevant strings, the weight is mostly captured by single $X$ Pauli matrices. The length distribution roughly matches the total distribution with a dilution factor $3^{-k}$ for length $k$. In the middle panel, we show then the time evolution of the three average lengths $\mathcal{L}(t)$, $\mathcal{L}_{\rm rel}(t;t)$ and $\mathcal{L}_{\rm dil}(t)$. We see that the relevant and diluted lengths are of the same order $\mathcal{O}(N^0)$. As for the total length, it grows quickly with time and saturates at around $\frac{3N}{4}$, which is the average length of an operator that would have equal weight in absolute value for all Pauli strings.

\subsubsection{System-size behaviour of the relevant string length at time $s$ for time $t>s$ \label{sec:systembehaviour}}

Let us now consider the relevant string length $\mathcal{L}_{\rm rel}(s;t)$ at time $s$ for time $t>s$. This requires to time-evolve the observable $O$ up to time $t$, then to keep in $O(t)$ only the $X$-strings, and then to time-evolve them back to time $s<t$. The operator $O_{\rm rel}(s;t)$ obtained this way is thus a linear combination of strings in the $X$ basis, evolved for a time $t-s$. As any generic operator evolved in time, we expect thus the average length of its strings $\frac{\tr[O_{\rm rel}(s;t) ^\dagger\ell(O_{\rm rel}(s;t))]}{\tr[O_{\rm rel}(s;t)^\dagger O_{\rm rel}(s;t)]}$ to grow with $t-s$, up to saturating at $\mathcal{O}(N)$. However, the relevant string length is \emph{not} this quantity. There are strings in $O_{\rm rel}(s;t)$ that do not appear in $O(s)$, and so that cannot contribute to noise. Taking the inner product of $O_{\rm rel}(s;t)$ with $\ell(O(s))$ the relevant string length excludes these strings. Defining the Pauli string decomposition
\begin{equation}
    O_{\rm rel}(s;t)=\sum_{P\in\mathcal{P}} c_P^{\rm rel}(s;t)P\,,
\end{equation}
 we have
 \begin{equation}
     \mathcal{L}_{\rm rel}(s;t)=\frac{\sum_{P\in\mathcal{P}} c_P^{\rm rel}(s;t) c_P(s) \ell_P}{\sum_{P\in\mathcal{P}} c_P^{\rm rel}(s;t) c_P(s)}\,.
 \end{equation}
We note that the denominator is $\sum_{P\in\mathcal{P}_x} |c_P(t)|^2$ which is positive. The coefficients $c_P(s;t)$ and $c_P(s)$ completely differ in general since $\sum_{P\in\mathcal{P}}|c_P(s)|^2=1$ whereas $\sum_{P\in\mathcal{P}}|c^{\rm rel}_P(s;t)|^2\ll 1$, and have generically different signs, which results in interference effects in these sums. 

As a consequence, since the number of strings of length $k$ grows exponentially with $k$, we expect stronger suppression of large string sectors due to interference. Therefore, \emph{if for both $O_{\rm rel}(s;t)$ and $O(s)$} the weight of short strings is not exponentially small, we expect $\mathcal{L}_{\rm rel}(s;t)=\mathcal{O}(N^0)$. If on the other hand one of them has only large strings, then we expect $\mathcal{L}_{\rm rel}(s;t)=\mathcal{O}(N)$. 

Let us again introduce a simple toy model to illustrate this behaviour as in Section \ref{revelantlengthtt}. We consider coefficients $c_P$ and $c_P^{\rm rel}$ that have all the same absolute values for $\ell_P\neq 1$, and introduce the free parameters $p_1,q_1$ which are the sum of absolute values squared of $c_P$ and $c_P^{\rm rel}$ respectively for $\ell_P=1$. Since the normalization of the coefficients cancels in the definition of the relevant string length, we will work with both $c_P$ and $c_P^{\rm rel}$ normalized to $1$. We also denote $p_k,q_k$ the sum of absolute values squared of $c_P$ and $c_P^{\rm rel}$ respectively for $\ell_P=k$, which have thus identical expressions in terms of $p_1,q_1$ as in \eqref{pk}. We now assume that the signs of $c_P,c_P^{\rm rel}$ are independent random variables taking values $\pm $ with probabilities $1/2$. Then according to the law of large numbers, we can write
\begin{equation}\label{gaussianpsik}
\begin{aligned}
\sum_{\substack{P\in\mathcal{P}\\ \ell_P=k}} c_P c_P^{\rm rel}&\sim \sqrt{\sum_{\substack{P\in\mathcal{P}\\ \ell_P=k}} (c_P c_P^{\rm rel})^2}\xi_k= \sqrt{\frac{p_kq_k}{3^k {N\choose k}}}\xi_k\,,
\end{aligned}
\end{equation}
with $\xi_k$ a Gaussian random variable with mean $0$ and variance $1$. In contrast, we have
\begin{equation}
\sum_{\substack{P\in\mathcal{P}\\ \ell_P=k}} |c_P| | c_P^{\rm rel}|= \sqrt{p_kq_k}\,.
\end{equation}
The interferences due to the signs bring thus a dilution factor $1/\sqrt{3^k {N\choose k}}$. From \eqref{pk} we have
\begin{equation}\label{gaussianpsik2}
\sum_{\substack{P\in\mathcal{P}\\ \ell_P=k}} c_P c_P^{\rm rel}\sim \frac{\sqrt{(1-p_1)(1-q_1)}}{1-3N/4^N} \frac{\sqrt{3^k {N \choose k}}}{4^N}\xi_k\,.
\end{equation}
Stirling's formula yields
\begin{equation}
    3^k {N\choose k}\sim \frac{e^{N(\alpha\log 3-\alpha\log \alpha-(1-\alpha)\log(1-\alpha))}}{\sqrt{2\pi N \alpha(1-\alpha)}}\,,
\end{equation}
with $\alpha=k/N$. The exponent is maximal for $\alpha=3/4$. Denoting $k=3N/4+x$ we have thus, summing \eqref{gaussianpsik2} over $k$
\begin{equation}
\begin{aligned}
    \sum_{P\in\mathcal{P}} c_P c_P^{\rm rel}&\sim \sqrt{\frac{p_1q_1}{3N}}\xi_1\\
    &+\frac{\sqrt{(1-p_1)(1-q_1)}}{(6\pi N)^{1/4}2^{N-1}} \sum_{x}\xi_{3N/4+x}e^{-\frac{4x^2}{3N}}\,.
\end{aligned}
\end{equation}
Similarly, we have
\begin{equation}
\begin{aligned}
    \sum_{P\in\mathcal{P}} c_P c_P^{\rm rel}\ell_P&\sim \sqrt{\frac{p_1q_1}{3N}}\xi_1\\
    &+\frac{3N}{4}\frac{\sqrt{(1-p_1)(1-q_1)}}{(6\pi N)^{1/4}2^{N-1}} \sum_{x}\xi_{3N/4+x}e^{-\frac{4x^2}{3N}}\,.
\end{aligned}
\end{equation}
The random variable $v=\sum_{x}\xi_{3N/4+x}e^{-\frac{4x^2}{3N}}$ has mean $\mathbb{E}[v]=0$ and variance
\begin{equation}
    \mathbb{E}[v^2]=\sum_{x} e^{-\frac{8x^2}{3N}}\sim \sqrt{\frac{3\pi N}{8}}\,.
\end{equation}
It follows
\begin{equation}
    \frac{1}{(6\pi N)^{1/4}2^{N-1}} \sum_{x}\xi_{3N/4+x}e^{-\frac{4x^2}{3N}}\sim \frac{\xi}{2^N} \,,
\end{equation}
with $\xi$ a Gaussian random variable with mean $0$ and variance $1$. Hence, if $\sqrt{p_1q_1}\gtrapprox \frac{N^{3/2}}{2^N}$, we obtain $\mathcal{L}_{\rm rel}(s;t)=\mathcal{O}(1)$. If $\sqrt{p_1q_1}\lessapprox \frac{N^{1/2}}{2^N}$, we get $\mathcal{L}_{\rm rel}(s;t)=\mathcal{O}(N)$. In the absence of STE for either $O(s)$ or $O_{\rm rel}(s;t)$, assuming uniform string weight we would have either $p_1=\frac{3N}{4^N}$ or $q_1=\frac{3N}{4^N}$, and so $\sqrt{p_1q_1}\leq \sqrt{3N}/2^N$, because $p_1,q_1\leq 1$. Hence the relevant length would be $\mathcal{O}(N)$. If on the contrary there is a STE for both $O(s)$ and $O_{\rm rel}(s;t)$, it constrains $p_1$ and $q_1$ to be only polynomially small, so e.g. of order $1/N$. In that case we have a relevant length $\mathcal{O}(1)$.

In the Ising model case we consider, since the relevant string sector at time $t$ is only the $X$ strings, the operator $\mathcal{O}_{\rm rel}(s;t)$ is comprised of $X$ strings evolved in time. Because there is a STE for single Pauli matrices $X_j$, we deduce that the operator $O_{\rm rel}(s;t)$ also contains short strings with non-exponentially small weight. We conclude that we expect the relevant length $\mathcal{L}_{\rm rel}(s;t)$ to also remain of order $\mathcal{O}(N^0)$ for $s<t$. 

In the right panel of Figure \ref{systemsizerelevant}, we provide a numerical check of this claim. We observe that the relevant length $\mathcal{L}_{\rm rel}(s;t)$ for $s<t$ remains bounded with $N$, of order $\mathcal{O}(N^0)$, whereas the total length $\mathcal{L}(s)$ is of order $\mathcal{O}(N)$. In order to check the importance of the signs in the coefficients $c_P(s)$ and $c_P^{\rm rel}(s;t)$ in the observation of a relevant length $\mathcal{O}(N^0)$, as suggested by our toy model, we consider the quantity
 \begin{equation}
     \mathcal{L}_{\rm abs}(s;t)\equiv \frac{\sum_{P\in\mathcal{P}} |c_P^{\rm rel}(s;t)| |c_P(s)| \ell_P}{\sum_{P\in\mathcal{P}} |c_P^{\rm rel}(s;t)| |c_P(s)|}\,.
 \end{equation}
 According to our arguments, this quantity should always be $\mathcal{O}(N)$ because of the absence of interference effects due to signs that dilute large string sectors. In the right panel of Figure \ref{systemsizerelevant}, we see indeed that $\mathcal{L}_{\rm abs}(s;t)$ computed numerically displays a $\mathcal{O}(N)$ behaviour, confirming our theory.\\

Together with the result of Section \ref{relation_relevant_noise} on the relation between the sensitivity to errors $\rho$ and the relevant length $\mathcal{L}_{\rm rel}(s;t)$, these results explain the low sensitivity to errors $\rho=\mathcal{O}(1/N)$ observed in this model.

\subsubsection{Toy model for the relation between $\rho$ and $\mathcal{L}_{\rm rel}(s;t)$ \label{toymodelrelation}}
In this subsection we come back on the relation between the perturbation of the expectation value and the relevant string length $\mathcal{L}_{\rm rel}(s;t)$, namely relation \eqref{rholrel}. We are going to derive that relation using a similar toy model as in Section \ref{sec:systembehaviour}. 

For an observable $O(t)=\sum_{P\in\mathcal{P}}c_P(t) P$, the expectation value in $|+...+\rangle$  is $\langle O(t)\rangle=\sum_{P\in\mathcal{P}_x}c_P(t)$. Let us assume again that the signs of $c_P(t)$ for $P\in\mathcal{P}_x$ a string of $X$ matrices are independent random variables with $\pm$ occurring with probability $1/2$. Then, we have
\begin{equation}
    \langle O(t)\rangle\sim \xi\sum_{P\in\mathcal{P}_x}|c_P(t)|^2\,,
\end{equation}
with $\xi$ a Gaussian random variable with mean $0$ and variance $1$. The right-hand side is exactly the Frobenius norm of $\Pi_X(O(t))$, which is conserved through unitary evolution. Since by definition $\Pi_X(O(t))=U_{t,s}O_{\rm rel}(s;t)U_{t,s}^\dagger$ for any $s\leq t$, we thus have for any $s\leq t$
\begin{equation}\label{eqotxi}
    \langle O(t)\rangle\sim \xi\sum_{P\in\mathcal{P}}|c^{\rm rel}_P(s;t)|^2\,.
\end{equation}
Let us now denote $\tilde{O}(t)$ the same observable in presence of noise, with an unspecified noise model for the moment. At leading order in the noise amplitude, the signs of the coefficients of $\tilde{O}(t)$ are unchanged, and so the same equation as \eqref{eqotxi} holds with the coefficients of $\tilde{O}(t)$. In this toy model, in analogy with \eqref{eq_new_noise_model} we define the dilution factor $\rho$ by
\begin{equation}\label{rhotoymodel}
    \frac{ \langle \tilde{O}(t)\rangle}{ \langle O(t)\rangle}=e^{-\epsilon \rho Nt}\,,
\end{equation}
where $\epsilon$ is the probability of error. We now fix $s\leq t$ and denote $\tilde{O}^s(t)$ the observable with noise inserted \emph{only} at time $s$, as well as $\tilde{c}^{\rm rel}_P(s;t)$ the coefficients of $\tilde{O}^s(s)$ projected onto the relevant space $S_{\rm rel}(s;t)$ at time $s$ for time $t$. Writing $\tilde{c}^{\rm rel}_P(s;t)=c^{\rm rel}_P(s;t)+\delta c^{\rm rel}_P(s;t)$, we have thus at leading order in perturbation
\begin{equation}
    \frac{ \langle \tilde{O}^s(t)\rangle}{ \langle O(t)\rangle}= 1+2\frac{\sum_{P\in\mathcal{P}}c_P^{\rm rel}(s;t)\delta c^{\rm rel}_P(s;t)}{\sum_{P\in\mathcal{P}}|c_P^{\rm rel}(s;t)|^2}+\mathcal{O}(\epsilon^2)\,.
\end{equation}
Since the inner product of $O_{\rm rel}(s;t)$ with an operator orthogonal to the relevant space $S_{\rm rel}(s;t)$ is $0$, we have
\begin{equation}
    \frac{ \langle \tilde{O}^s(t)\rangle}{ \langle O(t)\rangle}= 1+2\frac{\sum_{P\in\mathcal{P}}c_P^{\rm rel}(s;t)\delta c_P(s)}{\sum_{P\in\mathcal{P}}c_P^{\rm rel}(s;t)c_P(s)}+\mathcal{O}(\epsilon^2)\,,
\end{equation}
with $\delta c_P(s)$ the perturbation of the coefficients of $\tilde{O}^s(s)$ compared to $O(s)$. In case of a perturbation given by a depolarizing channel on each qubit with probability of error $\epsilon$, we have $\delta c_P(s)=-\frac{4}{3}\epsilon \ell_P c_P(s)$. Hence we exactly have
\begin{equation}
    \frac{ \langle \tilde{O}^s(t)\rangle}{ \langle O(t)\rangle}= 1-\frac{8\epsilon}{3} \mathcal{L}_{\rm rel}(s;t)+\mathcal{O}(\epsilon^2)\,.
\end{equation}
Inserting now noise at any time $s\leq t$, it follows that the coefficient $\rho$ in \eqref{rhotoymodel} is indeed related to the relevant string length as
\begin{equation}
    \rho=\frac{8}{3Nt}\sum_{s=1}^t \mathcal{L}_{\rm rel}(s;t)\,.
\end{equation}

\section{Area of Validity}
\label{sec_area_of_validity}

\begin{figure*}
\begin{center}

\includegraphics[width=\textwidth]{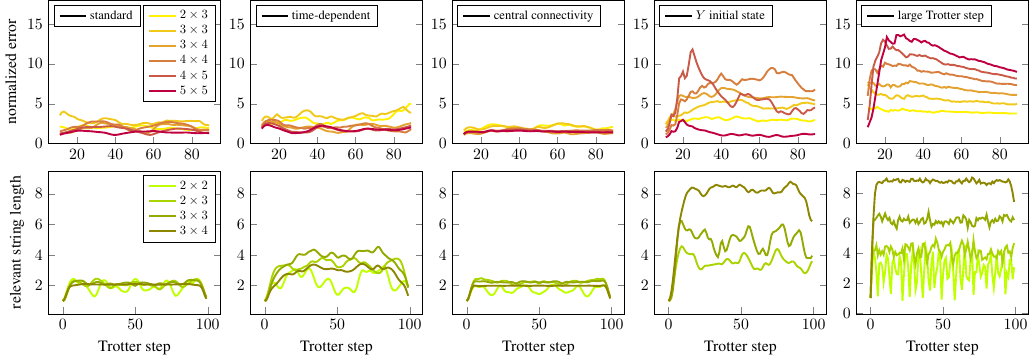}

\caption{\emph{Top panels:} Normalized difference between noisy and noiseless expectation value $\Delta O(t)$ as a function of Trotter step $t$, for different system sizes, in the standard setup described at the beginning of Section \ref{sec_area_of_validity} (``standard"), and in four variations: with $h(t)=\cos(4\pi t/100)$ with $t$ the Trotter step number (``time-dependent"), with additional $Z_1Z_j$ terms in the Hamiltonian for all $j=2,...,N$ (``central connectivity"), with initial state where $Y_j=+1$ for all $j$ (``Y initial state"), with Trotter step ${\rm d}t=1$ (``large Trotter step"), all computed with $1000$ noisy trajectories. \emph{Bottom panels:} Relevant string length $\mathcal{L}_{\rm rel}(s;t)$ as a function of $s$ for $t=100$, for different system sizes, in the same setup as the top panel directly above each bottom panel.}
\label {validityfigure}
\end{center}
\end {figure*}

We have provided evidence for the failure of a simple light cone theory of errors in section~\ref{sec_experimental_evidence}, as well as developed a microscopic theory in section~\ref{sec_microscopics}. We are now interested in the area of validity of our new error model, and in particular, which assumptions need to be broken to show a system-size independent errors for $k=\mathcal{O}(1)$-local observables. To this end, we will show numerical data for variations around a ``standard" setup, in which we test the influence of parameters on dilution of errors, comparing the prediction of our microscopic theory to numerical data. 

This ``standard" setup consists of the 2D Ising model implemented in Section \ref{sec_experimental_evidence}, i.e. Trotter circuit given by \eqref{Utrotter}, but with now magnetic field $h=1$ in the $X$ direction, Trotter step ${\rm d}t=0.1$, initial state $|+...+\rangle$ and measuring $O=S_x=\frac{1}{N}\sum_{j=1}^NX_j$. The noise model is taken to be a depolarizing channel applied after every Trotter step on every qubit
\begin{equation}
    \mathcal{N}(\rho)=\left(1-\frac{3\varepsilon}{4}\right)\rho+\frac{\varepsilon}{4}(X_j\rho X_j+Y_j\rho Y_j+Z_j\rho Z_j)\,,
\end{equation}
with $\varepsilon=0.001$. As a measure of distance between the noiseless expectation value $\langle O(t)\rangle$ and the noisy expectation value $\langle O(t)\rangle_{\rm noisy}$, we define the normalized difference
\begin{equation}
    \Delta O(t)=\frac{1}{\varepsilon t}\frac{\sum_{s=t-\Delta t/2}^{t+\Delta t/2}|\langle O(s)\rangle_{\rm noisy}-\langle O(s)\rangle|}{\sum_{s=t-\Delta t/2}^{t+\Delta t/2}|\langle O(s)\rangle|}\,,
\end{equation}
where $\Delta t$ is a time window taken to be equal to $\Delta t=20$. The smoothing over this time window allows one to avoid problematic cases where $\langle O(s)\rangle$ would vanish for some time $s$ because of oscillations around $0$. If the noisy expectation value is approximately $\langle O(t)\rangle_{\rm noisy}\approx e^{-\varepsilon\lambda t} \langle O(t)\rangle$, then this normalized difference provides the decay rate $\Delta O(t)\approx \lambda$ when the error rate $\varepsilon$ is small.

\subsection{Time-dependent Hamiltonians}
\label{sec_time_dep}
The microscopic explanation for the dilution of errors presented in Section \ref{sec_microscopics} in the case of Ising model Hamiltonian simulation also applies to the time-dependent case. Let us consider a time-dependent magnetic field in the $X$ direction, namely define $U(t)$ as
\begin{align}
    U(t) = \left( \prod_{\braket{ij}} e^{-i \mathrm{d}t Z_i Z_j} \right) \left( \prod_{i=1}^N e^{-i \mathrm{d}t \, h(t) X_i} \right)\,,
\end{align}
for some function $h(t)$. Then, the STE \eqref{conservationlaw} for the coefficients $c_X$ and $c_{ZZ}$ of the Pauli strings in $O(t)$ becomes, neglecting Trotter error,
\begin{equation}
\label{conservationlaw2}
    c_X(t+1)h(t+1)+2c_{ZZ}(t+1)=c_X(t)h(t+1)+2c_{ZZ}(t)\,.
\end{equation}
Hence, starting from $c_X(0)=1/N$, for generic $h(t)$ we would always have a non-zero weight for at least $c_X(t)$ or $c_{ZZ}(t)$. This means that at all time steps the operator $O(t)$ contains short strings with non-exponentially small weight, and similarly for $O_{\rm rel}(s;t)$. One notes that, however, contrary to the constant Hamiltonian case, this weight transfer can be blocked for some fine-tuned cases if at a step $t$, both $h(t+1)$ and $c_{ZZ}(t)$ turn out to be very small. In the generic case, the theory of Section \ref{sec_microscopics} applies and one expects dilution of error. We present numerical evidence for this in Figure \ref{validityfigure} with the label ``time-dependent". We see that the dilution of error is observed in this setting as well, in agreement with our microscopic theory.

\subsection{Connectivity of the Hamiltonian}
The Hamiltonian of the 2D Ising model we considered in Section \ref{sec_experimental_evidence} couples neighbouring qubits on a two-dimensional lattice and comprises only low-length Pauli strings. According to the microscopic explanation presented in Section \ref{sec_microscopics}, the local connectivity of the Hamiltonian is not required for the dilution of error. In order to provide numerical check of this claim, we modify the lattice by connecting all the qubits to the first qubit $j=1$ in the $ZZ$ term of \eqref{Utrotter}, the connections between other pairs of qubits remaining unchanged (a wheel $\&$ spokes geometry). This change maintains a number of couplings $\mathcal{O}(N)$. However, we keep \emph{unchanged} the noise model, with a depolarizing channel applied on every qubit after every Trotter step. On actual quantum hardware, since the $j=1$ qubit is acted on $\mathcal{O}(N)$ times per Trotter step, noise would differ and the $j=1$ qubit would receive $\mathcal{O}(N)$ times more noise than on the regular 2D lattice. We choose here to keep unchanged the noise model to isolate the effect of qubit connectivity on dilution of error. This model removes in particular any potential light-cone argument as light-cones quickly cover the entire system through the first qubit to which all other qubits are connected. The numerical results are presented in the ``central connectivity" panel of Figure \ref{validityfigure}. We see that the dilution of error is observed in this setting as well, in agreement with our microscopic theory.

\subsection{Sensitivity to initial state}
Dilution of error is however sensitive to the initial state. The initial state crucially determines the relevant space $S_{\rm rel}(s;t)$ and the projection of the observable onto it $O_{\rm rel}(s;t)$. Let us change for example the initial state to be a product state with $Y_j=+1$ at each site
\begin{equation}
|\psi\rangle=\otimes_{j=1}^N \frac{1}{\sqrt{2}}\left(\begin{matrix}
        1\\ i
    \end{matrix}\right)\,.
\end{equation}
In that case, the relevant space is
\begin{equation}
    S_{\rm rel}(s;t)={\rm span}\{ U^\dagger_{t,s}P U_{t,s}\,,\quad P\in \mathcal{P}_y\}\,,
\end{equation}
with $\mathcal{P}_y$ the set of all $Y$ strings. Equation \eqref{conservationlaw} still holds in that case, because we are only changing the initial state, and not the Hamiltonian or the observable measured. This ensures that there are always some short strings with non-negligible weight in $O(t)$. However, the relevant subspace is now obtained by time-evolving $Y$ strings only. And crucially, $Y$ strings do not have any trace overlap with the Hamiltonian $H$. The equation \eqref{conservationlaw} for $O=Y$ (when performing the backward evolution entering the definition of the relevant subspace) becomes in that case
\begin{equation}
    c_X(t)h+2c_{ZZ}(t)=0\,,
\end{equation}
which does not prevent $c_X$ or $c_{ZZ}$ from becoming arbitrarily small, and so does not guarantee non-zero weight for short strings. The same holds true for $H^2$, which has no trace overlap with $Y$. Non-zero trace overlaps occur only at third order $H^3$. As a consequence, while the string length distribution of $O(s)$ is unchanged since it is independent of the initial state, the relevant operator $O_{\rm rel}(s;t)$ is completely modified, and the weight of short strings will be spread onto strings of length up to $6$ through the STE obtained for $H^3$. The dilution of error should be much weaker in this case, and even maybe barely visible for the system sizes that can be simulated classically. We present numerical confirmation on this in Figure \ref{validityfigure}.

\subsection{Sensitivity to Trotter step size}
We finally comment on the sensitivity of dilution of error to Trotter step size. The STE \eqref{conservationlaw} requires the absence of Trotter error to hold, i.e. holds up to corrections $\mathcal{O}({\rm d}t)$. In case of a non-negligible Trotter step size ${\rm d}t$, a similar equation would hold with respect to the Floquet Hamiltonian $H_F(\mathrm{d}t)$ defined as
\begin{align}\label{UFloquet}
    \exp\left(-i \mathrm{d}t H_F\right) \equiv \left( \prod_{\braket{ij}} e^{-i \mathrm{d}t  Z_i Z_j} \right) \left( \prod_{i=1}^N e^{-i \mathrm{d}t \, h X_i} \right)\,.
\end{align}
This Hamiltonian admits an expansion in $\mathrm{d}t$ whose first terms are
\begin{equation}
    \begin{aligned}
        H_F(\mathrm{d}t)=&H_X+H_{ZZ}-\frac{i\mathrm{d}t}{2}[H_{ZZ},H_{X}]\\
        &+\frac{ \mathrm{d}t^2}{12}([H_{X},[H_{ZZ},H_{X}]]-[H_{ZZ},[H_{ZZ},H_{X}]])\\
        &+\mathcal{O}({\rm d}t^3)\,,
    \end{aligned}
\end{equation}
with $H_X=h\sum_{j=1}^N X_j$ and $H_{ZZ}=\sum_{\langle i,j\rangle}Z_iZ_j$. Importantly, using the Baker-Campbell-Hausdorff formula, one has that the length of the Pauli strings appearing at order ${\rm d}t^n$ grows linearly with $n$. As time increases, the initial weight $c_X(t=0)=1/N$ in front of Pauli strings $X_j$ will spread onto these Pauli strings through the STE for $H_F({\rm d}t)$.  As a consequence, larger ${\rm d}t$ should result in weaker dilution of error. In particular, if ${\rm d}t$ is large enough for the Floquet Hamiltonian to become scrambling, then the average string length in $H_F({\rm d}t)$ is $\mathcal{O}(N)$ and no dilution of error is expected. We show numerical evidence for the absence of dilution of error in Figure \ref{validityfigure} when the Trotter step is too large.\\

To summarize and conclude this section, we studied how modifying single parameters in a ``standard" Hamiltonian simulation setup preserves or destroys the dilution of error. We compared the numerical simulations to the prediction of our microscopic theory presented in Section \ref{sec_microscopics}, and obtained agreement. In particular, we saw that the relevant length and the effect of noise on expectation values scale similarly with system size in all the setups considered.

\section{Error Mitigation}
\label{sec_error_mitigation}
We have derived microscopic expressions for noisy observables in Section~\ref{sec_microscopics}. Equipped with these expressions, we use this section to develop two new error mitigation techniques that are particularly efficient in cases where errors dilute ($\rho \ll 1$) and act independently cf.~\eqref{eq_independent_errors} below.

\subsection{Recall: Probabilistic Error Cancellation}

Our first error mitigation technique can be seen as a simplification of Probabilistic Error Cancellation (PEC) \cite{temme2017error} or Zero Noise Extrapolation with Probabilistic Error Amplification (ZNE-PEA) \cite{kim2023evidence}, which we will briefly review here.
The basic mechanism behind PEC is that given Kraus operators $K$ such that the effect of the noise is given by the quantum channel
\begin{equation}
    \mathcal{N}(\rho)=\sum_{u=1}^{\kappa}K_u\rho K_u^\dagger\,,
\end{equation}
with $\kappa$ an integer and $\sum_{u=1}^\kappa K_u^\dagger K_u=I$, there exist real numbers $p_u$ with $\sum_u p_u=1$ such that
\begin{equation}\label{PEC}
  \sum_{u=1}^\kappa p_u K_u \mathcal{N}(\rho) K_u^\dagger =  \rho\,.
\end{equation}
If the $p_u$'s were positive, the left-hand side could be implemented on the hardware by applying $K_u$ on the state with probability $p_u$, exactly cancelling the effect of the noise channel $\mathcal{N}$ (provided we neglect errors occurring when applying $K_u$'s). However, in general, the $p_u$'s can be positive and negative. To implement the left-hand side of \eqref{noisechannel} probabilistically, one thus would have to apply each $K_u$ with probability $|p_u|\lambda$ where $\lambda=\frac{1}{\sum_u |p_u|}\leq 1$ (not to be confused with the noise rate introduced in section~\ref{sec_experimental_evidence}), and to multiply the result by $\sign(p_u)$. But then, instead of measuring the noiseless expectation value $\langle O(t)\rangle$, one actually measures $\lambda \langle O(t)\rangle$. The consequence of this attenuation factor $\lambda$ is to increase the number of shots required to reach a given precision. One needs to include $1/\lambda^2$ more shots to reach a given precision with PEC, compared to the case where there would be no noise on the hardware. In a circuit with $V$ quantum gates where the noise channel \eqref{noisechannel} is applied, the overhead cost of PEC in terms of shots is thus $\lambda^{-2V}$. For a simulation of a local Hamiltonian, one has typically $V=\mathcal{O}(Nt)$ with $N$ the system size and $t$ the number of Trotter steps. For local Kraus operators we have $\log \lambda=\mathcal{O}(\varepsilon)$. Hence the cost of this mitigation technique is $e^{\mathcal{O}(\varepsilon Nt)}$. Since $\varepsilon V$ is the typical number of errors occurring in the circuit, we obtain that PEC can only correct $\mathcal{O}(1)$ errors without prohibitive overhead in terms of shots.

\subsection{Perturbative mitigation: \texttt{LIN}}
\label{sec_LIN}
Let us come back to Eq \eqref{decomposition} that expresses the noisy value $\langle O(t)\rangle_{\rm noisy}$ as a power series in the probability of error $\varepsilon$. At order $\varepsilon$, this is
\begin{equation}
    \langle O(t)\rangle_{\rm noisy}=\langle O(t)\rangle+\varepsilon \langle\Sigma_1(t)\rangle +\mathcal{O}(\varepsilon^2)\,.
\end{equation}
Now, the quantity $\Sigma_1(t)$ can be measured on the hardware, by first measuring $D_1$ by inserting the Kraus operators $K_u$ inside the circuit, and then subtracting $D_0$, both defined in \eqref{definitiond}. This means we can form the operator
\begin{equation}
    O^{(1)}(t)=O(t)-\varepsilon\Sigma_1(t)\,,
\end{equation}
such that its measured noisy expectation value is
\begin{equation}
    \langle O^{(1)}(t)\rangle_{\rm noisy}=\langle O(t)\rangle +\mathcal{O}(\varepsilon^2)\,.
\end{equation}
This removes the term of order $\varepsilon$ from the expansion \eqref{decomposition}. This can be seen as applying PEC at first order in $\varepsilon$. We will refer to this noise mitigation procedure as \texttt{LIN}, for linear
\begin{equation}
   \langle O(t)\rangle_{\rm LIN }\equiv \langle O(t)\rangle_{\rm noisy}-\varepsilon\langle \Sigma_1(t) \rangle_{\rm noisy}\,.
\end{equation}
To measure $\Sigma_1$, one has to measure separately $D_0$ (the noisy expectation value) and $D_1$, the noisy expectation value with $1$ error inserted by hand at all possible places. The variance obtained on the mitigated value $\langle O(t)\rangle_{\rm LIN }$ is 
\begin{equation}
    \Delta_{\rm LIN}=\frac{\sigma_0^2}{S_0}(1+N_{\rm gate}\varepsilon)^2+\frac{\sigma_1^2}{S_1}(N_{\rm gate}\varepsilon)^2\,,
\end{equation}
where $S_{0,1}$ are the numbers of shots to measure $D_{0,1}$ the sectors with exactly zero and one errors, $\sigma_0^2$ the variance in measuring $D_0$, $\sigma_1^2$ the variance in measuring $D_1/N_{\rm gate}$, and $N_{\rm gate}$ the number of noisy gates in the circuit. At fixed shot budget $S=S_0+S_1$, the optimal shot distribution $\beta=\frac{S_0}{S_1}$ is
\begin{equation}
    \beta=\frac{\sigma_0}{\sigma_1}\frac{1+N_{\rm gate}\varepsilon}{N_{\rm gate}\varepsilon}\,.
\end{equation} 
We note that $N_{\rm gate}\varepsilon$ is the average number of errors in the circuit. For example when $\sigma_0=\sigma_1$, we see that we always have $S_0/S>1/2$, so the overhead to implement \texttt{LIN} is at most to double the number of shots.

We note that this scheme can be in principle generalized to higher orders in $\varepsilon$.

\subsection{Exponential mitigation: \texttt{EXP}}
\label{sec_EXP}
We now propose another mitigation strategy. It relies on the empirical observation that oftentimes, we have
\begin{equation}
    \frac{\langle \Sigma_n\rangle}{\langle \Sigma_0\rangle}\approx \frac{1}{n!}\left(\frac{\langle \Sigma_1\rangle}{\langle \Sigma_0\rangle}\right)^n\,.
\end{equation}
This approximation, loosely speaking, assumes that different errors do not ``influence" each other. In terms of the dynamical correlations appearing in the $\Sigma$ expansion of Section \ref{sigmasection}, this means relations of the type
\begin{align}
    \braket{K_{(2)} K_{(1)} O K_{(1)}^\dagger K_{(2)}^\dagger} \braket{O} = \braket{K_{(1)} O K_{(1)}^\dagger} \braket{K_{(2)} O K_{(2)}^\dagger}
    \label{eq_independent_errors}\,.
\end{align}
We define the \texttt{EXP} noise mitigation formula as
\begin{equation}
   \langle O(t)\rangle_{\rm EXP }\equiv \langle O(t)\rangle_{\rm noisy}\exp\left(-\varepsilon\frac{\langle \Sigma_1(t)\rangle_{\rm noisy}}{\langle O(t)\rangle_{\rm noisy}}\right)\,.
\end{equation}
This requires to measure $\langle O(t)\rangle_{\rm noisy}$ and $\langle \Sigma_1(t)\rangle_{\rm noisy}$ on the quantum computer. It has thus exactly the same complexity as the \texttt{LIN} mitigation technique. However, we will see below that it achieves better precision in a number of cases.

The variance obtained on the mitigated value $\langle O(t)\rangle_{\rm EXP }$ is 
\begin{equation}
\begin{aligned}
    \Delta_{\rm EXP}=&\bigg[\frac{\sigma_0^2}{S_0}\left(1+\varepsilon \frac{\langle D_1\rangle_{\rm noisy}}{\langle O(t)\rangle_{\rm noisy}}\right)^2\\
    &\qquad\qquad\qquad+\frac{\sigma_1^2}{S_1}(N_{\rm gate}\varepsilon)^2\bigg]\\
    &\times\exp\left(-2\varepsilon\frac{\langle \Sigma_1(t)\rangle_{\rm noisy}}{\langle O(t)\rangle_{\rm noisy}}\right)\,.
\end{aligned}
\end{equation}
At fixed shot budget $S=S_0+S_1$, the optimal shot distribution $\beta=\frac{S_0}{S_1}$ is
\begin{equation}
  \beta=\frac{\sigma_0}{\sigma_1}\frac{\left|1+\varepsilon \frac{\langle D_1\rangle_{\rm noisy}}{\langle D_0\rangle_{\rm noisy}}\right|}{N_{\rm gate}\varepsilon}\,.
\end{equation}

\subsection{Impact of dilution of errors on the performance of  mitigation}
Let us evaluate the effect of dilution of errors on the \texttt{LIN} mitigation technique. In case of perfect dilution of errors, the terms in the $\Sigma$ expansion in \eqref{decomposition} scale as
\begin{equation}
    \langle \Sigma_n(t)\rangle=\mathcal{O}\left(\frac{t^n N^0}{n!}\right)\,,
\end{equation}
whereas without dilution of errors they scale as $\mathcal{O}(\frac{t^n N^n}{n!})$. In order for $\alpha^n/n!$ to be smaller than $\epsilon>0$ for some $\alpha>0$, one needs to have $n>e\alpha+\log 1/\epsilon$. Hence, in absence of dilution of error, one needs to remove all the sectors with less than $n=\mathcal{O}(\varepsilon tN)$ errors. Since the cost in removing a sector with $n$ errors is exponential in $n$, the mitigation cost is exponential in $\varepsilon tN$ in absence of dilution of errors.

In contrast, in presence of dilution of errors, one only needs to remove all the sectors with less than $n=\mathcal{O}(\varepsilon t)$ errors. The mitigation cost is thus exponential in $\varepsilon t$, but not in system size $N$.

\subsection{Hardware performance}
\begin{figure}
\begin{center}

\includegraphics[width=0.48\textwidth]{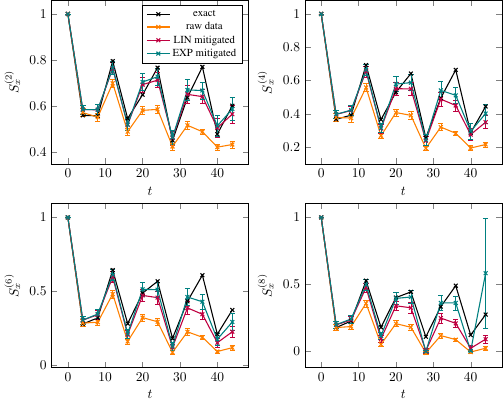}

\caption{\textbf{Experimental Data.} Same as the top panels of Figure \ref{fig_decayrates_hardware}, but for the observable $S_x^{(k)}$ defined in \eqref{eq_definition_Sxk} for $k=2,4,6,8$ in reading direction.}
\label {hardwareresultshigher}
\end{center}
\end {figure}

We now evaluate the performance of these noise mitigations techniques on the hardware data. Firstly, in the two top panels of Figure \ref{fig_decayrates_hardware} at the beginning of this manuscript, we plotted the measured magnetization $S_x^{(1)}$, comparing the exact value with the raw data and the mitigated data. While the raw data departs from the exact data because of hardware noise by several standard deviations, we see that both \texttt{LIN} and \texttt{EXP} mitigation techniques succeed in recovering the noiseless signal within one or two error bars. The very good performance of these mitigations techniques is particularly visible in the cumulated plot, where the mitigated expectation values almost superimpose with the exact values. For this observable we also see that \texttt{LIN} and \texttt{EXP} output almost the same values. In Figure \ref{hardwareresultshigher} we then perform the same comparison but with the more noisy observables $S_x^{(k)}$ for $k>1$. At larger values of $k$, we see a significant difference between \texttt{LIN} and \texttt{EXP}. While the \texttt{LIN} mitigation fails in recovering the noiseless signal at larger values of $k$, we see that the \texttt{EXP} mitigation keeps constant performance.

\subsection{Numerical tests}
\begin{figure}
\begin{center}

\includegraphics[width=0.5\textwidth]{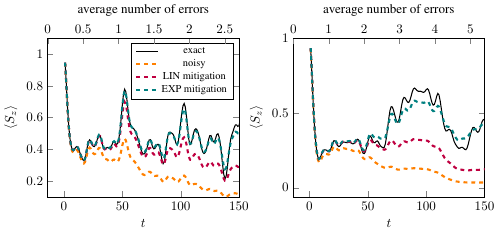}

\caption{Comparison of efficiency of \texttt{LIN} and \texttt{EXP} mitigation techniques, using exact density matrix simulations with depolarizing channel occurring on every qubit after every Trotter step with probability $0.0015$ (left panel) and $0.003$ (right panel), in Ising models on square lattice $3\times 4$, starting from $|0...0\rangle$, measuring $S_z=\frac{1}{N}\sum_{j=1}^N Z_j$, with ${\rm d}t=0.1$, and with $h_x=-1.8$  (left panel), $h_x(t)=-2 \cos(\omega t {\rm d}t)$ where $\omega=0.5$ (right panel).}
\label {expmitigationtechnique}
\end{center}
\end {figure}

In order to test further these noise mitigation techniques, we perform numerical simulations with exact evolution of density matrices. Despite being limited to system sizes $\lessapprox 12$, they allow for noisy simulations without error bars. In Figure \ref{expmitigationtechnique} we compare the efficiency of \texttt{LIN} and \texttt{EXP} on an Ising model simulation, both for a constant Hamiltonian case and for a time-dependent Hamiltonian. We observe that both methods recover very well the original signal in presence of an average number of errors $\lessapprox 1$. For larger number of errors, we observe on these plots that the \texttt{EXP} mitigation technique performs particularly well. Besides the cases plotted, we observed that the \texttt{EXP} mitigation tends to perform less well for adiabatic settings where parameters are slowly varied with time. However, in non-adiabatic settings, we observed excellent performance of \texttt{EXP} in many simulation cases, and the cases plotted are not fine-tuned.

\section{Exactly solvable case: free fermions}
\label{sec:freefermions}
While it is notoriously difficult to obtain provable bounds for general noisy quantum circuits, there is one situation, in which such bounds can be obtained analytically: the dynamics of the one-dimensional transverse-field Ising model. We prove the system size independence of the late-time behaviour of the $\Sigma$ expansion in this system and give analytic expressions for its value. Specifically, we consider the Hamiltonian
\begin{equation}\label{Ising}
    H=\sum_{j=1}^N Z_j Z_{j+1}+hX_j\,,
\end{equation}
with some magnetic field $h\geq 0$ and with periodic boundary conditions. We fix the initial state $|+...+\rangle$ and the observable $O=S_x$, and define the unitary operator that implements one Trotter step
\begin{equation}\label{1dtrotterstep}
    U=\exp\left(-i\mathrm{d}t\sum_{j=1}^N Z_jZ_{j+1} \right)\exp\left(-i\mathrm{d}t h\sum_{j=1}^N X_j \right)\,,
\end{equation}
for some Trotter step $\mathrm{d}t$. We consider two different noise models. The first noise model contains $X$ errors only, i.e. inserts a $X_j$ at position $j$ with some probability $\varepsilon$ after each Trotter step. We note that while the noiseless model is quadratic in fermions, the density matrix describing the state in presence of noise is \emph{not} Gaussian. In that case, we compute that the leading behaviour at large time of $\langle \Sigma_n(t)\rangle$ is
\begin{equation}
   \langle \Sigma_n(t)\rangle=\frac{(-\lambda t)^n}{n!}\langle \Sigma_0(t)\rangle+\mathcal{O}(t^{n-1}N^0)\,,
\end{equation}
with
\begin{equation}
    \lambda=4(1-m)\,,
\end{equation}
where $m$ is the expectation value of the magnetization at large times $\langle S_x(t)\rangle$. In particular, it is of order $N^0$ with respect to system size, and not $N^n$. In the case $\mathrm{d}t\to 0$ without Trotter error, we have
\begin{equation}\label{lambdaff}
    \lambda=\begin{cases}
        \frac{2}{h^2}\quad \text{for }h>1\\
       2\quad \text{for }h\leq 1\,.
    \end{cases}
\end{equation}
We detail the calculations leading to this result in the Appendix \ref{appendix_freefermion}. In Figure \ref{systemsizeindependenceff} we observe numerically for ${\rm d}t\to 0$ that $\lambda$ is indeed the decay rate of the observable $S_x$ at low error rate $\varepsilon$, namely that
\begin{equation}
    \langle S_x(t)\rangle_{\rm noisy}\sim e^{-\lambda \varepsilon t} \langle S_x(t)\rangle\,.
\end{equation}
Also in Appendix~\ref{appendix_freefermion}, we study another noise model given by a depolarizing channel with probability $\varepsilon$ applied on each site after each Trotter step. In that case, we show that $\langle \Sigma_1(t)\rangle=-\lambda t \langle \Sigma_0(t)\rangle+\mathcal{O}(t^0)$ with an explicit value for $\lambda=\mathcal{O}(N^0)$.

% \begin{equation}
%     \lambda=
% \end{equation}
% which is again of order $N^0$.

% compute $\langle \Sigma_1(t)\rangle$ and show that it is $\mathcal{O}(t N^0)$, again of order $N^0$ with respect to system size. Besides, we express the late-time decay rate of the system as the smallest imaginary part of a $N\times N$ matrix. This late-time decay rate is found to be $\mathcal{O}(N^0)$. We observe numerically that while the amplitude of $\frac{\langle \Sigma_1(t)\rangle}{t \langle \Sigma_0(t)\rangle}$ gives a good account of the attenuation of the signal at short or intermediate times, the late-time $t\to\infty$ decay rate \emph{differs} from this quantity.

\begin{figure}
\begin{center}

\includegraphics[width=0.48\textwidth]{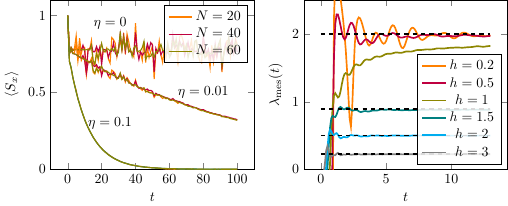}

\caption{\emph{Left panel:} Expectation value of $S_x$ as a function of time $t$ in the 1D Ising model, for different system sizes $N$ and different noise levels $\eta\equiv \varepsilon/{\rm d}t=0,0.01,0.1$ with ${\rm d}t\to 0$ (from top to bottom), for a magnetic field $h=1.5$, averaged over $1000$ realizations of noise. \emph{Right panel:} Decay rate $\lambda_{\rm mes}(t)=-\frac{1}{t}\log \frac{\langle S_x\rangle_{\rm noisy}}{\langle S_x\rangle}$ as a function of simulation time $t$, for different values of magnetic field $h$, in size $N=100$, with a noise rate $\varepsilon/{\rm d}t=0.1$ with ${\rm d}t\to 0$, averaged over $1000$ realizations of noise. The theoretical values \eqref{lambdaff} are indicated with black dotted lines.}
\label {systemsizeindependenceff}
\end{center}
\end {figure}

\section{Correspondence between Trotter error and noise}
\label{sec_trotter_error}
Up to now in this manuscript, both in the experimental setup of Section \ref{sec_experimental_evidence} and in the theory Section \ref{sec_microscopics}, we focused exclusively on errors coming from quantum hardware imperfections, i.e. gate noise. However, Hamiltonian simulation implemented on digital quantum computers through product formulas like \eqref{Utrotter} also comes with a so-called \emph{Trotter error} for any finite time step $\mathrm{d}t>0$. While textbook error estimations indicate that for a local Hamiltonian like the Ising model, the Trotter error with step $\mathrm{d}t$ scales as $\mathcal{O}(N\tau\mathrm{d}t)$ with $\tau$ the total simulation time (i.e., number of Trotter steps multiplied by ${\rm d}t$), a scaling independent of $\tau$ has been observed in practice and is now well understood \cite{childs2021theory}. However, in a number of cases, it has also been observed numerically that Trotter error can be \emph{system-size independent} as well \cite{heyl2019quantum}, similarly to error coming from noise studied in this manuscript. In this section, we show that Trotter error and hardware noise share some structural similarities for Hamiltonian simulation. The microscopic mechanism of dilution of error presented in Section \ref{sec_microscopics} could thus also apply to Trotter error.

Let us recall the effect of Trotterization on expectation values. Given two Hamiltonians $H_1,H_2$ and a Trotter step $\mathrm{d}t$, we have
\begin{equation}
\begin{aligned}
    &e^{-i \mathrm{d}t H_1} e^{-i \mathrm{d}t H_2}=e^{-i \mathrm{d}t H_F}\\
    \text{with }\quad &H_F=H_1+H_2+ -\frac{i\mathrm{d}t}{2}[H_1,H_2]\\
    &-\frac{\mathrm{d}t^2}{12}([H_1,[H_1,H_2]]-[H_2,[H_1,H_2]])+\mathcal{O}(\mathrm{d}t^3)\,.
\end{aligned}
\end{equation}
Let us denote the terms in $H_F-(H_1+H_2)$ as $\mathrm{d}t\sum_{q=1}^Q c_q K_q$ for some Pauli strings $K_q$ and real coefficients $c_q$ of order $\mathcal{O}(1)$ that depend on $\mathrm{d}t$. For Hamiltonians $H_1,H_2$ that are sums of $\mathcal{O}(N)$ terms such that every site is touched by $\mathcal{O}(1)$ terms, we have $Q=\mathcal{O}(N)$, and the coefficients $c_q$ are of order $\mathcal{O}(1)$. Denoting $|\psi(\tau)\rangle_{\mathrm{d}t}$ the Trotterized state with time step $\mathrm{d}t$ for total simulation time $\tau$ and initial state $|+\rangle$, and considering $O$ an observable that is a sum of Pauli strings of a finite length, we have
\begin{equation}
    {}_{\mathrm{d}t}\langle \psi(\tau)|O|\psi(\tau)\rangle_{\mathrm{d}t}=E_0+{\mathrm{d}t} E_1+ {\mathrm{d}t}^2 E_2+\mathcal{O}({\mathrm{d}t}^3)\,,
\end{equation}
with
\begin{equation}
    \begin{aligned}
       & E_0=\langle +|U^{-\tau}O U^\tau |+\rangle\\
        &E_1=2\Im\sum_{q=1}^Q c_q \int_0^\tau {\rm d}s \langle +|U^{-s}K_q U^{s-\tau}O U^\tau |+\rangle\\
        &E_2=-2\Re \sum_{q,q'=1}^Q c_q c_{q'} \\
        &\quad\times\int_0^\tau {\rm d}s\int_0^s{\rm d}s'\langle +|U^{-s}K_q U^{s-s'}K_{q'}U^{s'-\tau}O U^\tau |+\rangle\\
    &+2\Re\sum_{q,q'=1}^Q c_q c_{q'}\\
    &\times\int_0^\tau {\rm d}s\int_0^s{\rm d}s' \langle +|U^{-s}K_q U^{s-\tau}O U^{\tau-s'}K_{q'} U^{s'} |+\rangle\,.
    \end{aligned}
\end{equation}
We denoted here $U=e^{-i(H_1 + H_2)}$. Let us first focus on $E_1$. At first sight, it is of order $\mathcal{O}(\tau N )$ because of the time integral and $\mathcal{O}(N)$ different Trotter error operators $K_q$. Compared to $E_0$, most of these terms actually display an important difference. $E_0$ is exactly the expectation value of the observable $O$ in the state $U^\tau |+\rangle$. In contrast, in $E_1$, when $s>0$, the terms can be interpreted as the matrix element of $O$ between two different states $U^\tau |+\rangle$ and $U^{\tau-s}K_qU^{s} |+\rangle$. Generically, these matrix elements are expected to become small whenever $s \ll\tau$. For example, the matrix element of a local observable between two Haar-random states will be in average exponentially small in system size. As a consequence, the time integral in $E_1$ can be restricted to $s=\mathcal{O}(1)$. In that case, we obtain the approximation
\begin{equation}
    E_1\approx -i\sum_{q=1}^Q c_q \int_{0}^{\Delta\tau} {\rm d}s \langle +| [K_q(\tau-s),O(\tau)]|+\rangle\,,
\end{equation}
for some $\Delta \tau$ of order $\mathcal{O}(1)$. One sees that for local Hamiltonian simulation, a light-cone argument applies and $[K_q(\tau-s),O(\tau)]$ will be only of order $\mathcal{O}(1/N)$ because $K_q$ is localized at some site, and $O$ and $K_q$ are separated by only a time $\Delta \tau$. With these approximations, $E_1$ is seen to be of order $\mathcal{O}(\tau^0 N^0)$.

Let us now consider the term $E_2$. The term is a priori of order $\mathcal{O}(N^2 \tau^2)$. However, the terms in the first line are expectation values of $O$ only when $q=q',s=s'$. According to our argument above, this restricts $q=q'$ and $s-s'$ to be of order $\mathcal{O}(1)$. This is the fundamental assumption in this Trotter-error-noise correspondence. Similarly, the terms in the second line can be written as expectation values of $O$ within the state $U^{\tau-s}K_qU^s|+\rangle$ only when $q=q',s=s'$. Otherwise, they are matrix elements of $O$ between two different states. This also restrains $q=q'$ and $s-s'$ to be of order $\mathcal{O}(1)$. Neglecting the terms that are not of this form yields the approximation that, for some $\Delta \tau=\mathcal{O}(1)$,
\begin{equation}
\begin{aligned}
    E_2\approx 2\Re\sum_{q=1}^Q c_q^2 \int_{0}^\tau {\rm d}s \int_{s-\Delta \tau}^s {\rm d}s' &\langle +|[K_q(s),O(\tau)]K_{q}(s')|+\rangle\,.
\end{aligned}
\end{equation}
Further assuming that the integrand is close to the value when $s=s'$, we obtain
\begin{equation}
    E_2\approx 2\Delta \tau\sum_{q=1}^Q c_q^2 \int_{0}^\tau {\rm d}s  \langle +|[K_q(s),O(\tau)]K_{q}(s)|+\rangle\,.
\end{equation}
In this approximation, the term $E_2$ in the Trotter error has the same structure as hardware noise, with Kraus operators $K_q$ inserted at the same time points. The Trotter error operators $K_q$ play the role of the Kraus operators in the noise channel, with coefficients $2\Delta\tau c_q^2$. If the system displays dilution of errors for gate noise as explained in Section \ref{sec_microscopics}, then it will also display dilution of Trotter error, and $E_2$ will scale only as $\mathcal{O}(\tau N^0)$.

\begin{figure}
\begin{center}
\includegraphics[scale=0.22]{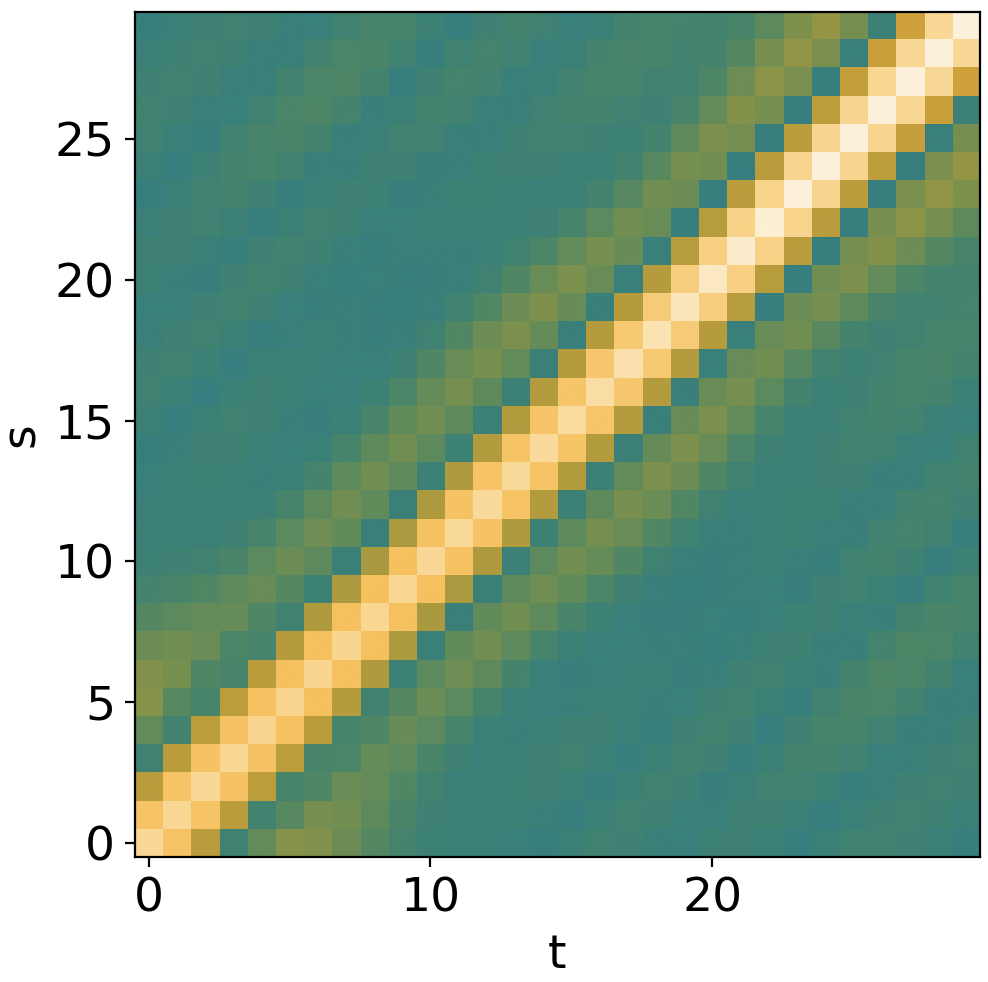}
\includegraphics[scale=0.22]{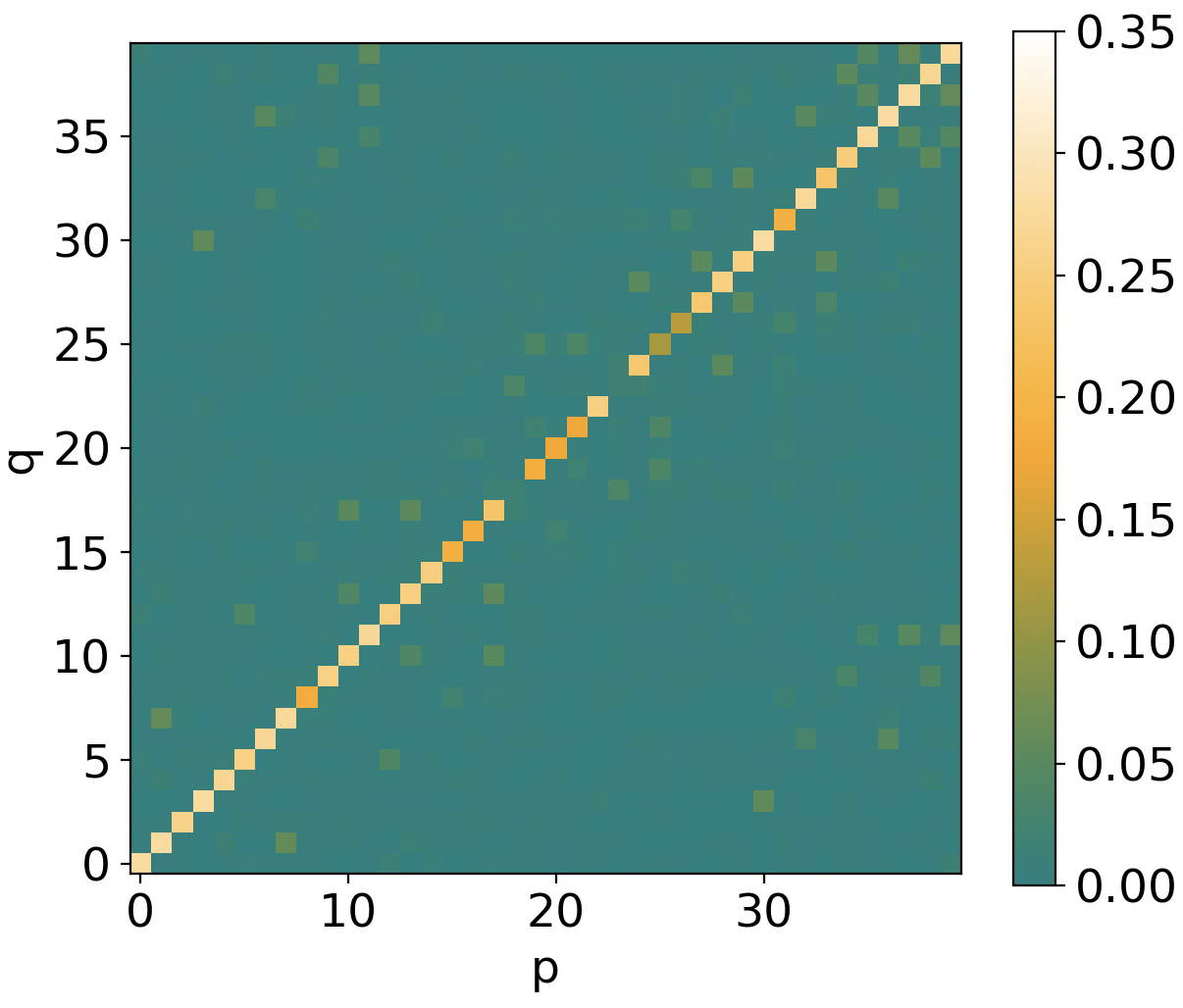}
\caption{\emph{Left panel:} Absolute value of the real part of $\langle +|K(s)O(t_*)K(t)|+\rangle$ with $t_*=30$, $O=X_j$ with $j$ at the center of the lattice, $K=Z_0Y_1$, in the standard setup of Section \ref{sec_area_of_validity}, as a function of $t$ and $s$, for system size $4\times 5$. \emph{Right panel:} same as left panel, but for $\langle +|K_q(s)O(t_*)K_{p}(s)|+\rangle$ with $s=15$, as a function of lattice bonds $p,q$ where $K$ is located, for an arbitrary ordering of bonds.}
\label {trotternoise_fig}
\end{center}
\end {figure}

We provide in Figure \ref{trotternoise_fig} a numerical check of the main assumptions to establish this correspondence between Trotter error and noise, which is the fact that $\langle +|K_q(s)O(t_*)K_p(t)|+\rangle$ can be neglected whenever $q\neq p$ or $|s-t|\gg 1$. We consider the 2D Ising model standard setup of Section \ref{sec_area_of_validity}, with observable $O=X_j$ at fixed $j$, and Kraus operator $K=Z_iY_j$ that arises on every bond from Trotter error in this model. We plot the real part of $\langle +|K_q(s)O(t_*)K_p(t)|+\rangle$ at fixed total simulation time $t_*$, for different times $s,t$ and different positions of the Kraus operators $q,p$. We see that, as claimed, this value is strongly suppressed whenever $q\neq p$ or $s-t$ away from $0$.

\section{Conclusion}
\label{sec_conclusion}
We have shown that the effect of quantum computer hardware noise on expectation values of local observables in Hamiltonian simulation setups can be much weaker than expected. In particular, in several settings noisy expectation values depart from the exact values by $\mathcal{O}(\varepsilon tN^0)$ for local Hamiltonian simulation, instead of $\mathcal{O}(\varepsilon tN)$ where $\varepsilon$ is the gate error probability, $t$ the number of Trotter steps and $N$ the system size. This \emph{dilution of error} goes beyond light-cone arguments and general arguments based on operator growth. We provided a microscopic explanation of this effect by making a connection  with the \emph{relevant string length} of the time-evolved operator. This number, loosely speaking, measures the average length of strings that belong to the subspace of strings of dimension $2^N$ (in the total space of operator of dimension $4^N$) that can contribute to the expectation value at time $t>s$. This microscopic mechanism is able to explain the sensitivity of dilution of error to parameters of the problem, such as time-dependence of the Hamiltonian or initial state.

If we depart from a standard noise model defined by some circuit volume, then the relative difficulty of quantum versus classical simulations is less clear than elaborated in \cite{kechedzhi2024effective}. In particular, we have seen that there exist families of circuits and observables for which the signal-to-noise ratio scales only as $\mathcal{O}(N^0)$ with system size $N$. In this case, the error mitigation techniques we have developed here provide a way to obtain accurate estimates of said observables with a number of shots that scales polynomially with system size, assuming a fixed, finite gate fidelity. These situations appear in generic quenches e.g., of strongly correlated non-integrable Ising models which---to the best of our knowledge---require classical resources that are exponential in system size.

We expect these findings to have large impact on problems requiring the measurement of local observables on $N \approx 100$ qubits with dense circuits containing $\approx 10^4$ two-qubit gates, the regime also considered in e.g.~\cite{meglio2024challenges}. A quantum computer with effective two-qubit gate errors of $10^{-3}$ \cite{decross2024computational} would produce output states with global fidelities less than $10^{-4}$, a signal that would require taking billions of shots to extract useful information. The situation is very different for the simulations of local observables for which dilution of errors applies: For example, our hardware data implies a sensitivity to errors $\rho \approx 3.5/N$ for two-point correlation functions $\langle \sum_{i,j} X_i X_j \rangle$ in the 2D Ising model quench of Fig.~\ref{fig_decayrates_hardware}. This value of $\rho$ implies a signal $\braket{O}_\mathrm{noisy}/\braket{O}_\mathrm{noiseless} > 70\%$ for these ``$100 \times 100$"-circuits. Assuming an intrinsic variance of the observable $\sigma^2 = 0.1$, the remaining relative bias can be suppressed to less than $10\%$ with standard error less than $0.1$ using 5000 shots and the LIN mitigation technique proposed in section~\ref{sec_LIN} (this is a conservative estimate in the sense that the EXP technique (cf. section~\ref{sec_EXP}) would likely perform better). Assuming shot times of one second, classically intractable problems could therefore be solved with a noisy quantum computer in less than two hours. Thus, providing further evidence for or against the dilution of errors, as well as the factorisation of errors~\eqref{eq_independent_errors} becomes an important objective, for which current quantum hardware can be employed.

\section*{Acknowledgements}
We acknowledge helpful comments with Michael Foss-Feig and Eli Chertkov, as well as initial discussions with them sparking the interest in this topic. We thank Yuta Kikuchi for useful comments on the draft. We acknowledge Karl Mayer for providing us with the error probabilities of Quantinuum's H1-1 hardware. E.G. acknowledges support by the Bavarian Ministry of Economic Affairs, Regional Development and Energy (StMWi) under project Bench-QC (DIK0425/01). H.D. acknowledges support by the German Federal Ministry of Education and Research (BMBF) through the project
EQUAHUMO (grant number 13N16069) within the funding
program quantum technologies - from basic research to market. The experimental data reported in this work was produced by the Quantinuum H1-1 trapped ion quantum computer, Powered by Honeywell, in March 2024.

\appendix

\section{Details on hardware circuit implementation}
\label{appendix-experiemnt}

\subsection{Setting}
\begin{figure}[H]
\begin{center}

\includegraphics[width=0.25\textwidth]{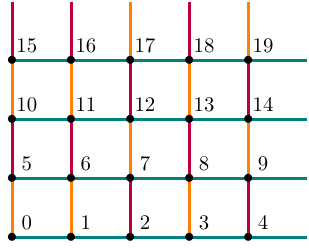}

\caption{Schematic representation of the Ising model implemented on the hardware. $ZZ$ rotations on sites separated by an orange edge are applied first, then purple edges and then teal edges.} 
\label {order}
\end{center}
\end {figure}

We consider the Ising model on a square lattice of size $N=5\times 4$ with periodic boundary conditions, with a magnetic field $h=3$, and with the system initialized in the product state $|+...+\rangle$. Using a first-order Trotterization with a Trotter step ${\rm d} t=0.15$, we prepare the following wave function on the quantum computer for a number of steps $t$
\begin{equation}
    |\psi(t)\rangle=\left(\prod_{\langle j,k\rangle} e^{-i{\rm d} t Z_jZ_k}\prod_{j=1}^N e^{-ih{\rm d} t X_j}\right)^t|+...+\rangle\,.
\end{equation}
In order to improve parallelization on Quantinuum's H1-1 hardware, we apply the $X$ rotations in the order $j=0,...,19$, and the $ZZ$ rotations in the order $(j,k)=(0,1)$, $(2,3)$, $(4,0)$, $(6,7),...$, then $(1,2)$, $(3,4)$, $(5,6)$, $(7,8)$, $(9,5),...$ and then $(0,5)$, $(1,6)$, $(2,7)$, $(3,8)$, $(4,9)$, $(5,10),...$, see Figure \ref{order}. We observed indeed using Quantinuum's H1-1E emulator that the choice of gate ordering has a noticeable impact on the machine performance. We measure then the expectation value of $S_x$ and $S_x^{(k)}$ as defined in \eqref{eq_definition_Sxk}.

\subsection{Implementing the mitigation}
In order to implement the noise mitigation technique, one requires to know the Kraus operators describing the noise in a specific hardware. These have been well characterized for the H1-1 device. The single-qubit gate infidelity is $4\cdot 10^{-5}$. Given that we have $Nt\leq 800$ single-qubit gates in our circuit, we will neglect these errors and consider the single-qubit gates as perfect. As for the two-qubit gate $e^{i\frac{\theta}{2} Z_jZ_k}$, the Kraus operators describing the noise are Pauli matrices whose probabilities of occurrence are given in Table \ref{Kraushardware} for $\theta=\pi/2$. For $\theta<\pi/2$, the probabilities $p(\theta)$ are scaled according to
\begin{equation}
    p(\theta)=(0.418\theta+0.34)p(\pi/2)\,.
\end{equation}
These values were measured in November 2023 \cite{karl}. In our case, with $\theta=0.3$, this gives a reduction factor $0.47$ compared to Table \ref{Kraushardware}.

\begin{figure}[H]
\begin{center}
    \begin{tabular}{|c|c||c|c|}
\hline
Pauli error & probability& Pauli error & probability\\
\hline
   IX  & $0.000124$ &YX  & $4.7\cdot 10^{-6}$ \\
   IY  & $0.000124$ & YY  & $4.7\cdot 10^{-6}$ \\
   IZ  & $0.000327$ & YZ  & $0.000114$\\
   XI  & $0.000114$ &ZI  & $0.000221$  \\
   XX  & $4.7\cdot 10^{-6}$ & ZX  & $0.000124$ \\
   XY  & $4.7\cdot 10^{-6}$ & ZY  & $0.000124$\\
   XZ  & $0.000114$ &ZZ  & $0.000122$ \\
   YI  & $0.000114$ &&\\
\hline
\end{tabular}
\end{center}
    \caption{Probability of Pauli errors after a gate $e^{i\frac{\pi}{4}ZZ}$ on H1-1 devices, measured in June 2023 \cite{karl}.}
    \label{Kraushardware}
\end{figure}

Given the low error probability $<10^{-5}$ for double bit flip errors $XX$, $XY$, $YX$ and $YY$, we neglect them in our experiment. We are left with $11$ Pauli errors to measure on each edge. Due to the translation invariance of the model and of the observable, all the vertical (resp. horizontal) edges are equivalent to $(0,1)$ (resp. $(0,5)$). The gate ordering breaks this equivalence as an error occurring on e.g. sites $(0,1)$ or $(6,7)$ will be followed and preceded by a different number of $ZZ$ rotations. However, we observed on the emulator that the position of the horizontal or vertical error makes no noticeable difference. We thus fix the errors to always occur on either $(0,1)$ or $(0,5)$. Finally, at time $t$, the error can occur during each of the $t$ Trotter steps $1\leq s\leq t$.

Measuring the effect of each error requires only a small number of shots, as the measured amplitude are then multiplied by a small number (the error probability) which decreases the error bar. One thus obtain in principle $2\times 11\times t$ circuits to run to mitigate one single value of magnetization $S_x(t)$, each with a small number of shots. Due to a constraint of the H1-1 device where each circuit comes with a compilation overhead cost, it is problematic to run a large number of circuits with a small number of shots each. To go around this cost, we randomize the location in time $s$ of the error, as well as whether it is on a horizontal edge or vertical edge. At the beginning of the circuit, we measure in the $Z$ basis $\lceil\log_2(t)\rceil$ qubits prepared in $|+\rangle$, and interpret the measurement outcomes as the binary decomposition of  $s$. If this binary decomposition is larger than $t$, the shot is discarded. By measuring another qubit we draw randomly the location of the error on a horizontal or vertical edge. This technique reduces to $11$ the number of circuits to run per Trotter step $t$ to mitigate one value of magnetization.

\section{Ising chain in a transverse field}
\label{appendix_freefermion}

In this Appendix we analytically derive the dilution of errors for Hamiltonian evolution with the 1D Ising chain \eqref{Ising}.

\subsection{Recall: Diagonalization}

As is well-known, the 1D Ising chain \eqref{Ising} can be solved through a Jordan-Wigner transformation of the Pauli matrices into fermions $c_j$ \cite{pfeuty1970one,calabrese2012quantum}
\begin{equation}
    X_j=1-2c_j^\dagger c_j\,,\qquad Z_j=(c_j+c_j^\dagger)\prod_{m<j}(1-2c_m^\dagger c_m)\,,
\end{equation}
where $c_j$ satisfy $\{c_i,c_j\}=\{c^\dagger_i,c^\dagger_j\}=0$ and $\{c_i,c^\dagger_j\}=\delta_{i,j}$. The parity of the number of particles $\sum_{j=1}^N c_j^\dagger c_j$ is conserved by $H$ in \eqref{Ising}, so the system splits into even and odd number of particles sectors. The fermions are given periodic/anti-periodic boundary conditions $c_{L+1}=\pm c_1$ in the even/odd sector. We then introduce the Fourier components
\begin{equation}
    a_k=\frac{1}{\sqrt{N}}\sum_{j=1}^N e^{-ijk}c_j\,,
\end{equation}
with $k\in P_e=\{\frac{2(n+1/2)\pi}{N}\,,\, n=-N/2,...,N/2-1\}$ if the number of particles is even, and $k\in P_o=\{\frac{2n\pi}{N}\,,\, n=-N/2,...,N/2-1\}$ if the number of particles is odd. Since the initial state $|+\rangle$ contains no fermions, it is in the even particle sector, and in the rest of the section we will consider $k\in P_e\equiv P$ only. We will then denote $P_e^+\equiv P^+$ the subset of $P_e$ with positive elements. In terms of these fermions we have
\begin{equation}
\begin{aligned}
    \sum_{j=1}^N X_j&=2\sum_{k\in P^+} 1-(a_k^\dagger a_k+a_{-k}^\dagger a_{-k})\\
    \sum_{j=1}^N Z_jZ_{j+1}&=\sum_{k\in P^+} 2\cos k(a_k^\dagger a_k+a_{-k}^\dagger a_{-k})\\
    &+\sum_{k\in P^+} 2i\sin k(a_k^\dagger a_{-k}^\dagger+a_ka_{-k})\,.
\end{aligned}
\end{equation}
Let us thus introduce the $4$-dimensional space spanned by $|0\rangle, a_k^\dagger a_{-k}^\dagger|0\rangle,a_k^\dagger |0\rangle,a_{-k}^\dagger|0\rangle$ in this order. The operator $U$ implementing one Trotter step in \eqref{1dtrotterstep} is block-diagonal in these sectors $U=\otimes_{k\in P^+} U_k$, with
\begin{equation}\label{U}
    U_k=\left(\begin{matrix}
        V_k&\begin{matrix} 0 \\ 0\end{matrix}&\begin{matrix} 0 \\ 0\end{matrix}\\
        \begin{matrix} 0 & 0\end{matrix}&e^{-2i{\rm d}t\cos k} &0\\ \begin{matrix} 0 & 0\end{matrix}&0 &e^{-2i{\rm d}t \cos k}\\
    \end{matrix}\right)\,,
\end{equation}
where
\begin{equation}
    V_k=\left[\exp \left(\begin{matrix}
        0&-2{\rm d}t\sin k\\ 2{\rm d}t\sin k&-4i{\rm d}t\cos k
    \end{matrix}\right) \right]\left(\begin{matrix}
        e^{-2i{\rm d}t h}&0\\ 0& e^{2i{\rm d}t h}
    \end{matrix}\right)\,.
\end{equation}
We will write this matrix as
\begin{equation}\label{vmat}
    V_k=W_k \left(\begin{matrix}
        e^{-i{\rm d}t\varepsilon^+_k}&0\\0& e^{-i{\rm d}t\varepsilon^-_k}
    \end{matrix} \right) W_k^\dagger\,,
\end{equation}
with $W_k$ a unitary matrix, that we write as
\begin{equation}
    W_k=\left(\begin{matrix}
        \cos\theta_k & i\sin\theta_k e^{-i\varphi_k}\\
        i\sin\theta_k e^{i\varphi_k}& \cos\theta_k
    \end{matrix}\right) \,,
\end{equation}
for some $\theta_k,\varphi_k$. In the limit of zero Trotter step ${\rm d}t\to 0$, see e.g. \cite{calabrese2012quantum}, we have
\begin{equation}
    \varepsilon^\pm_k=2\cos k\pm 2\sqrt{1+h^2-2h\cos k}\,,
\end{equation}
and
\begin{equation}\label{costhetak}
  \cos 2\theta_k=\frac{h-\cos k}{\sqrt{1+h^2-2h\cos k}}\,,\qquad \varphi_k=0\,.  
\end{equation}

\subsection{Expansion in $\Sigma$ for the magnetization with $X$ errors}
In this Section we fix the noise model to be a purely $X$ Pauli error channel, with noise channel applied after every Trotter step
\begin{equation}
\begin{aligned}
     &\mathcal{N}(\rho)=(1-N\varepsilon)\rho+\varepsilon \sum_{j=1}^N X_{j} \rho X_{j}^\dagger\,.
\end{aligned}
\end{equation}
Namely, the density matrix $\rho_t$ after $t$ noisy Trotter steps satisfies
\begin{equation}
    \rho_{t+1}=\mathcal{N}(U\rho_t U^\dagger)\,,
\end{equation}
with initial state $\rho_0=|+\rangle\langle+|$. We will compute the different terms in the expansion \eqref{decomposition} for the noisy value of the magnetization $\langle O(t)\rangle_{\rm noisy}=\tr[O \rho_t]$ with $O=S_x$.\\

It is well-known that time-evolution under the Ising Hamiltonian \eqref{Ising} preserves the Gaussianity of density matrices for any magnetic field $h$, due to $H$ being quadratic in the fermions. Namely, if $\rho$ is a Gaussian density matrix (i.e., a density matrix in which all correlations functions can be expressed in terms of the $2$-point functions $c_i^\dagger c_j$ and $c_i c_j$ with Wick's theorem), then $U \rho U^\dagger$ is also Gaussian. One can thus compute its time evolution by just keeping track of the $2$-point functions, which is only a $2N\times 2N$ matrix, whereas $\rho$ is a $2^N\times 2^N$ matrix. Similarly, $X_j \rho X_j$ is also Gaussian when $\rho$ is Gaussian, since one can write $iX_j=e^{i\tfrac{\pi}{2}X_j}$. However, sums of Gaussian density matrices are (generically) not Gaussian. The density matrix describing the state corresponding to the time evolution under the noisy process with random insertions of $X$ operators is thus \emph{not} Gaussian.

\subsubsection{Zero error sector $\Sigma_0$ (recall)}
The first term $\Sigma_0$ of the expansion \eqref{decomposition} is the noiseless expectation value of the magnetization $S_x$ and its computation is straightforward. We write $S_x=\frac{2}{N}\sum_{k\in P^+}S_k$ with
\begin{equation}
    S_k=\left(\begin{matrix}
        1&0&0&0\\
        0&-1&0&0\\
        0&0&0&0\\
        0&0&0&0
    \end{matrix}\right)_k=1-a_k^\dagger a_k-a_{-k}^\dagger a_{-k}\,.
\end{equation}
Then we have
\begin{equation}
    \langle S_x\rangle=\frac{2}{N}\sum_{k\in P^+} \langle +| U^t_k S_k U_k^{-t} |+\rangle\,.
\end{equation}
From an explicit calculation, we find
\begin{equation}
    \langle +| U_k^t S_k U^{-t}_k|+\rangle=\cos^2(2\theta_k)+\sin^2(2\theta_k)\cos(t{\rm d }t\Delta\epsilon_k)\,,
\end{equation}
where we introduced $\Delta\epsilon_k=\epsilon_k^+-\epsilon_k^-$. Hence we obtain
\begin{equation}\label{sxt}
    \langle S_x(t)\rangle=\frac{2}{N}\sum_{k\in P^+}\cos^2(2\theta_k)+\sin^2(2\theta_k)\cos(t{\rm d }t \Delta \epsilon_k)\,.
\end{equation}
The summand comprises a term constant with $t$, and an oscillatory term. In finite size, the oscillatory term is a sum of a finite number of cosines, and so keeps oscillating for all times. In the thermodynamic limit $N\to\infty$ however, this magnetization is
\begin{equation}
    \langle S_x(t)\rangle=m(h,{\rm d }t)+\frac{1}{\pi}\int_0^\pi \sin^2(2\theta_k)\cos(t{\rm d}t \Delta\epsilon_k)\D{k}\,,
\end{equation}
with the magnetization 
\begin{equation}
    m(h,{\rm d }t)=\frac{1}{\pi}\int_0^\pi \cos^2(2\theta_k)\D{k}\,,
\end{equation}
that depends on $h,{\rm d}t$ through $\theta_k$. In the zero Trotter step limit ${\rm d }t\to 0$, the magnetization takes the simple form
\begin{equation}\label{magh}
    m(h)=\begin{cases}
        \frac{1}{2}\,,\qquad \text{if }h<1\\
        1-\frac{1}{2h^2}\,,\qquad \text{if }h>1\,.
    \end{cases}
\end{equation}
Let us investigate the behaviour of the oscillatory integral at large times. With a change of variable $u=\frac{1}{4}\Delta\epsilon_k$ we have
\begin{equation}
\begin{aligned}
    &\int_0^\pi \sin^2(2\theta_k)\cos(t{\rm d}t\Delta\epsilon_k)\D{k}=\\
    &\frac{1}{2h^2}\int_{|1-h|}^{1+h}\frac{\sqrt{(u^2-(1-h)^2)((1+h)^2-u^2)}}{u}\cos(4ut{\rm d}t)\D{u} \,.
\end{aligned}
\end{equation}
The function in the integrand multiplying $\cos(4ut{\rm d}t)$ vanishes at the ends of the interval $u=1+h,|1-h|$. However, the derivative diverges with a power $-1/2$. Hence according to standard results the magnetization converges as $\mathcal{O}(t^{-3/2})$ to its limit value at large $t$.

% The behaviour at large $t$ of this kind of oscillatory integral depends on the smoothness of the integrand $f(u)=\frac{\sqrt{(u^2-(1-h)^2)((1+h)^2-u^2)}}{u}$ in $u$ when seen as a periodic function of $u$ with the interval $[|1-h|,1+h]$ repeated on the entire real line. One sees that all the derivatives of $f(u)$ coincide at $1+h$ and $|1-h|$ since the function is invariant under changing $h\to -h$ when $u\in [|1-h|,1+h]$. Moreover, when $h\neq 1$, the function is regular. Hence when $h\neq 1$, this integral goes to $0$ exponentially fast in $t$. The magnetization converges exponentially fast to its limit value. At $h=1$ however, the integrand has a discontinuous derivative at $u=0$. This implies a power-law convergence $\sim 1/t^2$ of the magnetization to its limit value.

\subsubsection{Noisy evolution -- generalities}
We now consider the introduction of noise. Let us introduce the $2N$ Majorana fermions
\begin{equation}
    \eta_{2j}=c_j+c_j^\dagger\,,\qquad \eta_{2j+1}=i(c_j-c_j^\dagger)\,.
\end{equation}
They are hermitian and satisfy $\{\eta_j,\eta_k\}=2\delta_{j,k}$. The observable $S_x$ can be written in terms of them as
\begin{equation}
    S_x=\sum_{jk} \mathcal{S}_{jk} \eta_j\eta_k\,,
\end{equation}
with
\begin{equation}\label{Sab}
    \mathcal{S}_{a,b}=\frac{i}{2N}(\delta_{a_1,b_1+1}-\delta_{b_1,a_1+1})\delta_{a_0,b_0}\,,
\end{equation}
where we denote $a=2a_0+a_1$ with $a_1\in\{0,1\}$ and similarly for $b$. Given an operator $O$ that is quadratic in the Majorana fermions
\begin{equation}\label{Mjk}
    O=\sum_{jk} \mathcal{O}_{jk} \eta_j\eta_k\,,
\end{equation}
for some matrix $\mathcal{O}$, its time-evolution $UOU^\dagger$ is also quadratic in the fermions. Let us thus introduce the matrix $\mathcal{U}$ such that
\begin{equation}
    U \eta_j U^\dagger=\sum_{k} \mathcal{U}_{jk}\eta_k\,.
\end{equation}
Then we have $UOU^\dagger=\sum_{jk} \mathcal{O}'_{jk} \eta_j\eta_k$ with $\mathcal{O}'=\mathcal{U}\mathcal{O}\mathcal{U}^\dagger$. Given an operator $O$ quadratic in the fermions, let us now evaluate the effect of the noise operator $N_o$ defined as
\begin{equation}
    N_{\rm oi}(O)=\sum_{j=1}^N [X_j,O]X_j\,.
\end{equation}
We write $[X_j,O]X_j=\frac{1}{2}[[X_j,O],X_j]$ and use the expression $X_j=i\eta_{2j}\eta_{2j+1}$ and
\begin{equation}
\begin{aligned}
    &[[\eta_{2j}\eta_{2j+1},\eta_a\eta_b],\eta_{2j}\eta_{2j+1}]\\
    &=8\eta_{2j}\eta_{2j+1}(\delta_{2j+1,a}\delta_{2j,b}-\delta_{2j+1,b}\delta_{2j,a})\\
    &+2(\eta_a\eta_b-\eta_b\eta_a)(\delta_{a,2j}+\delta_{a,2j+1}+\delta_{b,2j}+\delta_{b,2j+1})\,.
\end{aligned}
\end{equation}
We obtain
\begin{equation}
    N_{\rm oi}(O)=\mathcal{O}'_{ab}\eta_a\eta_b\,,
\end{equation}
with
\begin{equation}
\mathcal{O}'_{ab}=-4\mathcal{O}_{ab}+4\mathcal{O}_{ab} \delta_{a_0,b_0}(\delta_{a_1,b_1+1}-\delta_{b_1,a_1+1})\,,
\end{equation}
with same notations as in \eqref{Sab}. In our case, because of translation invariance of the Hamiltonian and of the noise, the matrix $\mathcal{O}$ always satisfies $\mathcal{O}_{a,b}=\mathcal{O}_{a+2j,b+2j}$ for any $j$, where the indices are taken modulo $2N$. In that case we have exactly
\begin{equation}
    N_{\rm oi}(O)=-4O-4iN o S_x\,,
\end{equation}
where $o=2\mathcal{O}_{2j,2j+1}$, independent of $j$. We now note that we have $\langle 0| \eta_a \eta_b|0\rangle=0$ whenever $a,b$ are not of the form $2j,2j+1$ for the same $j$, and in that case we have $\langle 0| \eta_{2j} \eta_{2j+1}|0\rangle=-i$. It follows that when $\mathcal{O}_{aa}=0$ for all $a$, we have
\begin{equation}
    \langle O\rangle\equiv \langle 0|O|0\rangle=-2i \sum_{j=1}^N \mathcal{O}_{2j,2j+1}\,.
\end{equation}
Hence
\begin{equation}\label{evolequationoise}
    N_{\rm oi}(O)=-4O+4 \langle O\rangle S_x\,,
\end{equation}
where $\langle O\rangle $ denotes the expectation value in the $|0\rangle$ state.

\subsubsection{One error $\Sigma_1$}
Let us now determine the contribution of one error to the $\Sigma$ expansion in \eqref{decomposition}, namely $\Sigma_1$. It is defined by
\begin{equation}
    \Sigma_1(t)=\sum_{s=1}^t U^{t-s}N_{\rm oi}\left(U^sS_xU^{-s}\right)U^{s-t}\,.
\end{equation}
Using \eqref{evolequationoise}, we have straightforwardly
\begin{equation}
\begin{aligned}
    \Sigma_1(t)&=\sum_{s=1}^t -4 S_x(t)+4\langle S_x(s)\rangle S_x(t-s)\\
    &=-4tS_x(t)+4\sum_{s=1}^t\langle S_x(s)\rangle S_x(t-s)\,,
\end{aligned}
\end{equation}
where we recall the definition $S_x(t)=U^tS_x U^{-t}$. In particular, this is of order $N^0$ and not of order $N$ as a priori expected. Using the expression \eqref{sxt} for $\langle S_x(t)\rangle$ and the equilibrium magnetization \eqref{magh}, at large $t$ this is
\begin{equation}\label{asymptoticsigma2}
    \langle\Sigma_1(t)\rangle=-t\lambda \langle S_x(t)\rangle+o(t)\,,
\end{equation}
with the decay rate
\begin{equation}\label{decayrateleading}
    \lambda(h,{\rm d}t)=4(1-m(h,{\rm d}t))\,,
\end{equation}
where the magnetization $m(h,{\rm d}t)$ is in \eqref{magh}. In the case ${\rm d}t\to 0$, this is
\begin{equation}\label{decayresult}
    \lambda(h)=\begin{cases}
        2\,,\qquad \text{if }h<1\\
        \frac{2}{h^2}\,,\qquad \text{if }h>1\,.
    \end{cases}
\end{equation}
Let us determine the corrections to the leading order in time of \eqref{asymptoticsigma2}. We saw that the magnetization $\langle S_x(t)\rangle$ converges as $1/t^{3/2}$ to its limit value. Hence the corrections in $\Sigma_1$ to the behaviour \eqref{asymptoticsigma2} are of order $\mathcal{O}(1/\sqrt{t})$.

\subsubsection{Multiple errors $\Sigma_n$}
Let us now consider the higher order terms $\Sigma_n$. We have by definition
\begin{equation}
    \Sigma_n(t)=\sum_{s=1}^t U^{t-s}N_{\rm noi}(\Sigma_{n-1}(s))U^{s-t}\,.
\end{equation}
Using \eqref{evolequationoise}, we get
\begin{equation}\label{recsigman}
    \Sigma_n(t)=\sum_{s=1}^t -4 U^{t-s}\Sigma_{n-1}(s) U^{s-t}+4\langle \Sigma_{n-1}(s)\rangle S_x(t-s)\,.
\end{equation}
 We are going to derive the asymptotic scaling in $t$ of $\Sigma_n$ by recurrence on $n$. We note that the term $U^{t-s}\Sigma_{n-1}(s) U^{s-t}$ is different from $\Sigma_{n-1}(t)$, as in the former all the $n-1$ errors must have occurred before time $s$, whereas in the latter the errors can occur at any time. Let us thus define property $P_m$ as the fact that $\langle U^{t-s} \Sigma_m(s) U^{s-t}\rangle$ is $\frac{(-s\lambda)^m}{m!}\langle S_x(t)\rangle+\mathcal{O}(s^{m-1}N^0)$ at large $s,t$. Property $P_m$ is true for $m=0$, since $U^{t-s} \Sigma_0(s) U^{s-t}=S_x(t)$. Let us assume $P_m$ true for some $m$. Then from \eqref{recsigman} we have
 \begin{equation}
     \begin{aligned}
         U^{t-s} \Sigma_{m+1}(s) U^{s-t}=\sum_{u=1}^s& -4 U^{t-u}\Sigma_m(u) U^{u-t}\\
         &+4 \langle \Sigma_m(u)\rangle S_x(t-u)\,.
     \end{aligned}
 \end{equation}
 Using $P_m$, we have at large $t$
 \begin{equation}
 \begin{aligned}
     &\langle U^{t-s} \Sigma_{m+1}(s) U^{s-t}\rangle=\sum_{u=1}^s -4\frac{(-u\lambda)^m}{m!}\langle S_x(t)\rangle+\mathcal{O}(u^{m-1}N^0)\\
     &\qquad+4 \left(\frac{(-u\lambda)^m}{m!}\langle S_x(u)\rangle+\mathcal{O}(u^{m-1}N^0) \right)\\
     &\qquad\qquad\qquad\times\left(m(h)+\mathcal{O}((t-u)^{-3/2})\right)\,,
 \end{aligned}
 \end{equation}
with $m(h)$ the magnetization in \eqref{magh}. We obtain
\begin{equation}
    \langle U^{t-s} \Sigma_{m+1}(s) U^{s-t}\rangle=\frac{(-\lambda s)^{m+1}}{(m+1)!} \langle S_x(t)\rangle+\mathcal{O}(s^mN^0)\,,
\end{equation}
with the decay rate $\lambda$ defined in \eqref{decayresult}. Hence, property $P_m$ is true for all $m$ and we have thus the asymptotic scaling
\begin{equation}
    \langle  \Sigma_{n}(t) \rangle=\frac{(-\lambda t)^{n}}{n!} \langle S_x(t)\rangle+O(t^{n-1}N^0)\,.
\end{equation}
All the $\Sigma_n$'s are thus of order $t^nN^0$, and not $t^nN^n$ as a priori expected.

In general, the corrections to these asymptotic behaviour are indeed of order $t^{n-1}N^0$. The subleading contributions from $\Sigma_n$ at order $t$ will combine to provide a change of decay rate with $\varepsilon$, namely
\begin{equation}
\begin{aligned}
    \sum_{n=0}^\infty \varepsilon^n \langle \Sigma_n(t)\rangle=&\langle S_x(t)\rangle\\
    &-t(\lambda+\varepsilon\lambda^{(1)}+\varepsilon^2 \lambda^{(2)}+...)\langle S_x(t)\rangle\\
    &+\mathcal{O}(t^2)\,,
\end{aligned}
\end{equation}
with $\lambda^{(i)}$ coefficients. The decay rate of the observable is thus $\varepsilon\lambda+\varepsilon^2\lambda^{(1)}+\varepsilon^3 \lambda^{(2)}+...$ with a non-trivial dependence on $\varepsilon$.

\subsection{Decay rate with depolarizing noise}

\subsubsection{Noisy evolution -- generalities}
We now consider a depolarizing noise channel
\begin{equation}
    N_{\rm oi}(O)=\frac{1}{3}\sum_{j=1}^N \sum_{P\in \{X,Y,Z\}}[P_j,O]P_j\,.
\end{equation}
For simplicity we will consider only the continuous time evolution limit ${\rm d}t\to 0$.

Let us consider the effect of this depolarizing channel on a term $c_i^\dagger c_j$ with $i<j$. When written in terms of Pauli matrices, $c_i^\dagger c_j$ is a sum of $4$ strings of Pauli matrices all starting from site $i$ and going to site $j$, thus of size $|i-j|+1$. Since every Pauli matrix in $\{X,Y,Z\}$ commutes with itself and anticommutes with the others, we have
\begin{equation}
    [X_j,S]X_j+[Y_j,S]Y_j+[Z_j,S]Z_j=-4S\,,
\end{equation}
whenever $S$ is a Pauli string with a non-trivial Pauli matrix $\in\{X,Y,Z\}$ at site $j$, and otherwise, this quantity vanishes. It follows that we have for example for $i< j$
\begin{equation}
N_{\rm oi}(X_iZ_{i+1}...Z_{j-1}X_j)=-\frac{4}{3}(|i-j|+1)X_iZ_{i+1}...Z_{j-1}X_j\,.
\end{equation}
The application to this formula to $c_i^\dagger c_j$ comes with some subtlety. Although the model in terms of the Pauli matrices is translation invariant, the Jordan-Wigner transformation to map the Pauli matrices to the fermions is \emph{not} invariant by translation as it selects one site $j=0$ as the beginning of the strings. $Z_1Z_N$ is mapped to $(c_1^\dagger-c_1)(1-2c_2^\dagger c_2)...(1-2c_{N-1}^\dagger c_{N-1})(c_N^\dagger+c_N)$ which is not quadratic in the fermions. The quadraticity and translation invariance at the fermionic level are recovered by noting that within a fixed sector where $\prod_{j=1}^NX_j=+1$, we can always multiply the terms in the Hamiltonian by $\prod_{j=1}^NX_j$ without changing any expectation value. This multiplication removes then the string of $1-2c_j^\dagger c_j$ in $Z_1Z_N$. However, in terms of the Pauli matrices, the Hamiltonian contains now a long string $Y_1X_2...X_{N-1}Y_N$ which breaks translation invariance. While this multiplication is necessary to keep the model quadratic in the fermions and thus solvable, the original qubit Hamiltonian (which is the one that would be implemented on a quantum computer) does not have this long Pauli string term and is indeed translation invariance, and so should be the effect of noise. To reflect that translation invariance at the level of the fermions, we impose thus
\begin{equation}
N_{\rm oi}(c_i^\dagger c_j)=\varphi(i-j)c_i^\dagger c_j\,.
\end{equation}
where we defined the $N$-periodic function of $j$
\begin{equation}
    \varphi(j)=-\frac{4}{3}(|j|+1)\,,\qquad -N/2<j\leq N/2\,.
\end{equation}
We impose the same equation for terms like $c_ic_j$. As for $c_i^\dagger c_i$, it is equal to $\frac{I_i-X_i}{2}$, so we have
\begin{equation}
    N_{\rm oi}(c_i^\dagger c_i)=\frac{2}{3}-\frac{4}{3}c_i^\dagger c_i\,.
\end{equation}
With these choices, $c_1^\dagger c_N$ gets as much noise as $c_2^\dagger c_1$. We checked numerically that the simulation of the fermionic system with this noise model agrees with the application of $N_{\rm oi}$ on the system when described by qubits and Pauli matrices.

Let us now consider a traceless and translation-invariant observable $O$ that is quadratic in the fermions $c_i,c_i^\dagger$. We can always write it as
\begin{equation}
\begin{aligned}
   O=&\sum_{k\in P_+}A_k(1-a_k^\dagger a_k-a_{-k}^\dagger a_{-k})\\
    &+iB_k (a_k^\dagger a_{-k}^\dagger+ a_{k}a_{-k})\\
    &+C_k (a_k^\dagger a_{-k}^\dagger-a_{k}a_{-k})\,,
\end{aligned}
\end{equation}
in terms of the momentum space fermions $a_k$, with $A_k,B_k,C_k$ some real coefficients. Its time evolution under the Hamiltonian $H$ and depolarizing noise is
\begin{equation}
    \partial_t O=i[H,O]+\eta N_{\rm oi}(O)\,,
\end{equation}
for some $\eta>0$ which is the error rate per unit time. Let us denote $x_{k,a}$ for $k\in P_+$ and $a=1,2,3$ the vector containing the coefficients $x_{k,1}=A_k$, $x_{k,2}=B_k$, $x_{k,3}=C_k$. Then we can write the evolution equation on $O$ as a matrix evolution equation on $x$
\begin{equation}
    \partial_t x =\mathcal{M}x\,,
\end{equation}
with $\mathcal{M}=\mathcal{H}+\eta \mathcal{N}$. Here, $\mathcal{H}$ is block-diagonal in the $k$ space with elements that are $3\times 3$ matrices given by
\begin{equation}
    \mathcal{H}_{k,k}=\left(\begin{matrix}
        0&0&\sin(2\theta_k)\\
        0&0&-\cos(2\theta_k)\\
        -\sin(2\theta_k)&\cos(2\theta_k)&0
    \end{matrix}\right)\Delta\epsilon_k\,.
\end{equation}
As for $\mathcal{N}$, it is diagonal in the $a$ index, and takes values
\begin{equation}
\begin{aligned}
    \mathcal{N}_{k,q;1,1}&=\hat{\varphi}(k-q)+\hat{\varphi}(k+q)\\
    \mathcal{N}_{k,q;2,2}&=\hat{\varphi}(k-q)-\hat{\varphi}(k+q)\\
    \mathcal{N}_{k,q;3,3}&=\hat{\varphi}(k-q)-\hat{\varphi}(k+q)\,.
\end{aligned}
\end{equation}
where we defined the Fourier transform for $k\in P$
\begin{equation}
    \hat{\varphi}(k)=\frac{1}{N}\sum_{j=1}^N e^{ijk}\varphi(j)\,.
\end{equation}
The initial condition on $x(t)$ is
\begin{equation}
    x_{k,a}(t=0)=\frac{\delta_{a,1}}{N}\,.
\end{equation}
The expectation value of $S_x$ at time $t$ is then given by
\begin{equation}
    \langle S_x(t)\rangle=\sum_{k\in P_+}x_{k,1}(t)\,.
\end{equation}
The decay rate at large times is given by the smallest real part of the eigenvalues of the matrix $\mathcal{M}$.

\subsubsection{One error $\Sigma_1$}
We have the $1$ error contribution
\begin{equation}
    \Sigma_1(t)=\sum_{s=1}^t U^{t-s}N_{\rm oi}(U^sOU^{-s})U^{s-t}\,.
\end{equation}
Time evolution for time $s$ modifies the coefficients $A_k,B_k,C_k$ as
\begin{equation}
    \begin{aligned}
        A_k(s)&=[\cos^2(2\theta_k)+\sin^2(2\theta_k)\cos(s\Delta\epsilon_k)]A_k\\
        &\qquad+\sin(4\theta_k)\sin^2(s\Delta\epsilon_k/2)B_k\\
&\qquad-\sin(2\theta_k)\sin(s\Delta\epsilon_k)C_k\\
    \end{aligned}
\end{equation}
and
\begin{equation}
    \begin{aligned}
        B_k(s)&=\sin(4\theta_k)\sin^2(s\Delta\epsilon_k/2)A_k\\
        &\qquad+[\sin^2(2\theta_k)+\cos^2(2\theta_k)\cos(s\Delta\epsilon_k)]B_k\\
&\qquad+\cos(2\theta_k)\sin(s\Delta\epsilon_k)C_k\\
    \end{aligned}
\end{equation}
and
\begin{equation}
    \begin{aligned}
        C_k(s)&=\sin(2\theta_k)\sin(s\Delta\epsilon_k)A_k\\
&\qquad-\cos(2\theta_k)\sin(s\Delta\epsilon_k)B_k\\
&\qquad+\cos(s\Delta\epsilon_k)C_k\,.
    \end{aligned}
\end{equation}
We thus get at large times
\begin{equation}
\begin{aligned}
     \langle \Sigma_1(t)\rangle=&\frac{t}{N}\sum_{k,q\in P_+}\cos^2(2\theta_k)\hat{\varphi}(k-q)\cos^2(2\theta_q)\\
     &+\sin(2\theta_k)\cos(2\theta_k)\hat{\varphi}(k-q)\sin(2\theta_q)\cos(2\theta_q)\\
     +&\frac{t}{N}\sum_{k,q\in P_+}\cos^2(2\theta_k)\hat{\varphi}(k+q)\cos^2(2\theta_q)\\
     &-\sin(2\theta_k)\cos(2\theta_k)\hat{\varphi}(k+q)\sin(2\theta_q)\cos(2\theta_q)\\
     &+\mathcal{O}(t^0)\,.
\end{aligned}
\end{equation}
Fourier-transforming back, this is
\begin{equation}
     \langle \Sigma_1(t)\rangle=-t\lambda+\mathcal{O}(t^0)\,,
\end{equation}
with
\begin{equation}
    \lambda=-\frac{1}{2}\sum_{j=-N/2}^{N/2} \varphi(j)(|\alpha_j|^2+|\beta_j|^2)\,,
\end{equation}
where we introduced
\begin{equation}
    \begin{aligned}
        \alpha_j&=\frac{1}{N}\sum_{k\in P}\cos^2(2\theta_k)e^{ijk}\\
        \beta_j&=\frac{1}{N}\sum_{k\in P}\cos(2\theta_k)\sin(2\theta_k)e^{ijk}\,.
    \end{aligned}
\end{equation}
In the thermodynamic limit $N\to\infty$, these coefficients become
\begin{equation}
\begin{aligned}
    \alpha_j&=\int_{-\pi}^\pi \frac{(h-\cos k)^2}{1+h^2-2h\cos k} e^{ijk}\D{k}\\
    \beta_j&=\int_{-\pi}^\pi \frac{(h-\cos k)\sin k}{1+h^2-2h\cos k} e^{ijk}\D{k}\,.
\end{aligned}
\end{equation}
For $h\neq 1$, these are the Fourier coefficients of regular $2\pi$-periodic functions, so they decay exponentially fast with $j$. Hence the coefficient $\lambda$ is of order $\mathcal{O}(N^0)$, and not $\mathcal{O}(N)$ as a priori expected.

\end{document}